\documentclass[a4paper,fleqn,usenatbib,useAMS]{mnras}
\usepackage{graphicx}
\usepackage{natbib}
\usepackage{amsmath}
\usepackage{mathtools}
\usepackage{amssymb}
\usepackage{wasysym}
\usepackage{txfonts}
\usepackage[T1]{fontenc}
\usepackage{ae, aecompl}

\usepackage{latexsym,longtable,epsf,ulem}
\usepackage{cases} 
\usepackage{enumerate} 

\newcommand{\bnt}{\beta_{\rm nt}}

\newcommand{\mmska}{SKA-MPG telescope}

\title[CMB foreground with the \mmska]
{CMB foreground measurements through broad-band radio spectro-polarimetry:
prospects of the SKA-MPG telescope} 
\author[A. Basu et al.]{Aritra Basu,$^{1,2}$\thanks{E-mail:
aritra@physik.uni-bielefeld.de.} Dominik J. Schwarz,$^1$ Hans-Rainer
Kl{\"o}ckner,$^2$ Sebastian von \newauthor
Hausegger,$^3$ Michael Kramer,$^2$ Gundolf Wieching$^2$ and Blakesley Burkhart$^{4,5}$
	\\
$^1$Fakult{\"a}t f{\"u}r Physik, Universit{\"a}t Bielefeld, Postfach 100131, D-33501 Bielefeld, Germany\\
$^2$Max-Planck-Institut f{\"u}r Radioastronomie, Auf dem H{\"u}gel 69, D-53121 Bonn, Germany\\
$^3$Niels Bohr Institute \& Discovery Center, University of Copenhagen, Blegdamsvej 17, DK-2100 Copenhagen, Denmark\\
$^{4}$Center for Computational Astrophysics, Flatiron Institute, 162 Fifth Avenue, New York, NY 10010, USA \\
$^{5}$Department of Physics and Astronomy, Rutgers, The State University of New Jersey, 136 Frelinghuysen Rd, Piscataway, NJ 08854, USA \\
}

\begin{document}

\date{Accepted to be published in the MNRAS on 2019 June 11}

\pagerange{\pageref{firstpage}--\pageref{lastpage}} \pubyear{2002}

\maketitle

\label{firstpage}

\begin{abstract}

Precise measurement of the foreground synchrotron emission, which contaminates
the faint polarized cosmic microwave background radiation (CMB), is a major
challenge for the next-generation of CMB experiments. To address this,
dedicated foreground measurement experiments are being undertaken at radio
frequencies between 2 and 40 GHz. Foreground polarized synchrotron emission
measurements are particularly challenging, primarily due to the complicated
frequency dependence in the presence of Faraday rotation, and are best
recovered through broad fractional-bandwidth polarization measurements at
frequencies $\lesssim 5$ GHz. A unique opportunity for measuring the foreground
polarized synchrotron emission will be provided by the 15-m SKA-MPG telescope
operating in the frequency range 1.7 to 3.5~GHz (S-Band). Here, we present the
scope of a Southern sky survey in S-Band at 1 degree angular resolution and
explore its added advantage for application of powerful techniques, such as,
Stokes $Q$, $U$ fitting and RM-synthesis. A full Southern-sky polarization
survey with this telescope, when combined with other on-going efforts at
slightly higher frequencies, will provide an excellent frequency coverage for
modeling and extrapolating the foreground polarized synchrotron emission to CMB
frequencies ($\gtrsim80$~GHz) with rms brightness temperature better than 10~nK
per 1 degree$^2$. We find that this survey will be crucial for understanding
the effects of Faraday depolarization, especially in low Galactic latitude
regions. This will allow better foreground cleaning and thus will contribute
significantly in further improving component separation analyses and increase
usable sky area for cosmological analysis of the \textit{Planck} data, and the
\textit{LiteBIRD} mission in the future.

\end{abstract}

\begin{keywords}  
cosmic microwave background -- polarization -- methods : data analysis -- radio
continuum : galaxies, ISM 
\end{keywords}

\section{Introduction}

The cosmic microwave background (CMB) radiation is the first light observable
after the Big Bang and is richly encoded with information on fundamental
physics. Sensitive measurements of the CMB and its spatial anisotropies have
been the most successful probe in our understanding of the Universe
\citep{penzi65, smoot92, hinsh13, planck2018I}. On angular scales $> 1^\circ$,
the next frontier in CMB experiments lies in the precise measurement of
anisotropies in its polarization states, namely, the gradient-type
\textit{E}-mode and the curl-type \textit{B}-mode, for the detection of
primordial gravitational waves and imprints of reionization history of the
Universe. A measurement of the \textit{B}-mode signal from gravitational waves
in the angular power spectrum of the polarized CMB on angular scales $\gtrsim
10^\circ$ \citep{Polnarev1985} could pin down the energy scale of cosmological
inflation \citep{liddle1994, lyth97, Guo2011, Planck2018X}. Detection of
\textit{reionization bumps} on angular scales $\gtrsim40^\circ$, in both the
\textit{E}- and \textit{B}-modes, would give us information on the astrophysics
and dark matter physics of the epoch of reionization \citep{zelda97, kogut03}.

Anisotropy studies of the CMB, both in intensity and polarization, from current
space- and ground-based observations have revealed that the lack of proper
foreground Galactic emission models is a bottle-neck in inferring orders of
magnitude fainter cosmological signals. In total intensity, the foreground
Galactic continuum emission consists of: synchrotron emission from relativistic
electrons gyrating in ambient magnetic fields, bremsstrahlung (also referred to
as the free--free emission) from thermal electrons scattering off ions,
anomalous microwave emission (AME) and emission from thermal dust grains. In
polarization, synchrotron and thermal dust emissions are the major
contributors. Because each of these foreground components dominate in different
frequency regimes, dedicated sky surveys are necessary to constrain them at
their respective frequencies. Once constrained, they are then extrapolated to
60--250~GHz frequency range, where the relative contribution from the CMB is
significant, and are subtracted.  Hence, in order to measure CMB anisotropies
with high precision, accurate understanding of the spectrum of each of the
foreground emission components is imperative for subsequent extrapolation and
subtraction at CMB frequencies.

The synchrotron and the free--free emissions dominate over the other foreground
components at low radio frequencies and are therefore best constrained below
10~GHz. For a physical description of total and polarized synchrotron emission
from the Milky Way, broad-bandwidth observations with multi-frequency coverage
are necessary. Due to the lack of suitable large sky-area surveys below 10~GHz,
current analyses typically rely on a mix of fits using power-laws, template
maps and simulated spectra. Modelling of the total and polarized synchrotron
spectrum is limited to minimal, empirical parametrization using the
\citet{hasla82} all-sky survey at 408 MHz as a template, while the free--free
emission is constrained by using ancillary Galactic H$\alpha$ maps
\citep{dicki03, planck2015XXV, planck2018IV}. To better constrain the polarized
and total synchrotron emission, current efforts are targeting frequencies below
10~GHz, for example, the recent S-Band Polarization All Sky Survey (S-PASS) at
2.3~GHz \citep{krach18, carre19} and the on-going C-Band All Sky Survey
(C-BASS) at 5~GHz \citep{jones18}. 

Extrapolating the polarized synchrotron foreground to CMB frequencies from
measurements at low radio frequencies ($\lesssim 10$ GHz) is especially
challenging in the presence of Faraday rotation. This is because, depending on
the properties of a magneto-ionic medium and how the synchrotron emitting and
Faraday rotating plasmas are mixed, the polarized synchrotron emission varies
in a complicated way, both with frequency and spatially. Imprints of
complicated Galactic Faraday rotating medium on the polarized synchrotron
emission spectrum is robustly captured through broad-bandwidth
spectro-polarimetric observations, especially in the frequency range 2 to
10~GHz, and can be modelled with analytical descriptions of turbulent
magneto-ionic media.

The upcoming \mmska, equipped with a sensitive receiver operating between 1.7
and 3.5~GHz (S-Band) provides a unique opportunity to perform a
broad-bandwidth, spectro-polarimetric survey of the entire Southern sky at
$1^{\circ}$ angular resolution. Although, {the \mmska\ was not
originally} designed for CMB foreground studies, in this paper we explore the
prospects of removal of foreground synchrotron and free--free emissions for a
spectro-polarimetric survey performed at S-Band {with it}. The
1.75~GHz wide, continuous frequency coverage will offer a deeper understanding
of the Milky Way's radio continuum spectrum and the opportunity to constrain
the Galactic total and polarized synchrotron emissions, along with the
free--free emission from radio continuum data alone on a pixel-by-pixel basis.

This work is organized as follows. We briefly summarize the current scientific
interest regarding CMB measurements in Section~\ref{sec:cmb}. In
Section~\ref{sec:syncsurvey} we briefly discuss the present understanding
regarding Galactic synchrotron emission from previous surveys. We present a
physical approach to describe the total radio continuum foreground spectrum and
discuss about effects of Faraday depolarization on polarized spectrum in
Section~\ref{sec:radio_fg}. In Section~\ref{sec:m2skad} we present the
specifications of the \mmska. In Section~\ref{sec:sband} we discuss the
advantages of adding S-Band data in estimation of foreground contributions at
CMB frequencies. In Section~\ref{sec:quijote-south} we point out the
possibility of a Southern-sky survey with the \mmska\ in the 12 to 18 GHz
frequency range and discuss our findings in Section~\ref{sec:summary}.

\section{CMB frontiers}  \label{sec:cmb}

Several decades of observational and theoretical efforts have established the
CMB to be the most precise and advanced probe of cosmology. The Planck 2018
results give an excellent summary of the state of the art \citep{planck2018I}.
The Universe can be described by the inflationary $\Lambda$ cold dark matter
(CDM) model, in which five cosmological parameters are measured at a level of
accuracy better than one per cent. In this minimal model, only the reionization
optical depth $\tau$ is constrained with a more modest accuracy of $10$ per
cent. Beyond the minimal model, upper limits on hypothetical new effects have
been established, perhaps most impressive are tight upper limits on the
curvature of the Universe and on the sum of the three neutrino masses.

A number of fundamental physics questions remain open and can be tackled by
future CMB experiments. Those questions include --- (1) is the cosmic structure
seeded by quantum fluctuations during an epoch of cosmological inflation, or is
there another mechanism at work?, (2) are there deviations from Einstein's
general relativity?, (3) what is the nature of dark matter and is the dark
energy a cosmological constant?, (4) what is the origin of the
matter-anti-matter asymmetry?, and (5) what are the neutrino masses? CMB also
holds clues on important astrophysical questions, such as, (6) what reionized
the Universe and how?, and (7) when did the first supermassive black holes form
and how is their formation related to galaxy formation?

Here we focus on the aspects that can be addressed on the largest angular
scales, as low-frequency foreground studies with single dish observations are
limited in angular resolution by dish size. Answers and/or contributions to
questions (1), (2), (3), and (6) definitely require or will benefit from
additional and improved information on large angular scales. The quest for the
physics of inflation will profit tremendously from a possible detection of (or
improved upper limits on) $B$-mode polarization caused by primordial
gravitational waves.  The history of reionization can be better constrained by
a precise measurement of the so-called reionization bumps in the $E$- and
$B$-mode angular power spectra.  Deviations from general relativity can be
constrained by probing primordial non-Gaussianity in the density fields on all
angular scales.  However, a detection of non-Gaussianity on larger angular
scales would make it simpler to discriminate potential astrophysical effects
that induce secondary non-Gaussian features on smaller angular scales. Hence,
deviations from general relativity can be constrained via cross-correlations of
the CMB temperature fluctuations with the large scale structure via the
integrated Sachs-Wolfe effect on large angular scales.

Another motivation to aim at a better understanding of all CMB foregrounds is
to disentangle several anomalies that have been observed in the CMB temperature
anisotropies on large angular scales as compared to that predicted by the
$\Lambda$-CDM model. These anomalies have been observed both with the
\textit{WMAP} and the \textit{Planck} satellites
\citep[see][]{Bennett_etal2011, Planck2015XVI, Schwarz_etal2016}, and therefore
instrumental or analysis pipeline effects are less likely. Statistical
significance of these anomalies is still inconclusive, i.e., they are seen at
most around $3\sigma$ significance when investigated one by one, but the
puzzling aspect is that there are several, seemingly unrelated and independent
anomalous aspects. Among these are a lack of variance and angular correlation
on the largest angular scales, alignment of phases of the lowest multipole
moments with the Solar dipole and among each other (quadrupole-octopole
alignment), hemispherical power asymmetry and power modulation and parity
asymmetry. At least some of those anomalies could be due to yet unidentified
features in the various CMB foregrounds. They will be further tested by CMB
polarization experiments and might point towards unexpected new physics, unless
foregrounds could eventually explain them.

To tackle these and other cosmological questions on large angular scales, the
next generation of space missions is currently in preparation. Most advanced in
the time line and funding effort is the Japan Aerospace Exploration Agency-led
\textit{LiteBIRD} mission \citep{suzuki18} and plans exist in Europe and the US
for a future large-scale CMB mission. Future ground-based CMB experiments will
target smaller angular scales and will put their focus on dedicated fields in
the sky. Many of them are on the Southern hemisphere, either at the South Pole
or in the Atacama desert at high altitude. Currently the most ambitious project
is CMB-S4 \citep{abaza16}. The success of these missions and experiments will
also depend on how well we understand CMB foregrounds, both in intensity and in
polarization.

\section{Existing observations of Galactic synchrotron polarization} \label{sec:syncsurvey}

To study the total Galactic synchrotron emission, several large sky-area
surveys have been undertaken in the past at frequencies below 5 GHz using
narrow frequency bandwidths \citep[e.g.,][to mention a few]{berkh72, hasla82,
reich86, uyani98, jonas98}. In polarization, most of the large sky-area surveys
have been performed around 1.4 GHz \citep[e.g.,][]{wolle06, testo08},
{and} only a few surveys have gone up to 5 GHz
\citep{sun07}.\footnote{For brief summary on previous large sky-area surveys
using narrow frequency bands, we refer the interested reader to \citet{jones18}
and \url{https://lambda.gsfc.nasa.gov/product/}.} Majority of these
polarization surveys were not designed to constrain CMB foregrounds and were
performed using narrow bandwidths at frequencies below 2 GHz. A combination of
inadequate sensitivity and sky-coverage makes them unsuitable to glean complete
information on the properties of the Galactic magneto-ionic medium and thereby
for CMB foreground studies. 

However, previous polarization surveys near 1.4~GHz have revealed that in the
region along the Galactic plane, roughly within Galactic latitude ($b$) $|b|
\lesssim30^\circ$, the polarized emission is significantly depolarized
\citep{wolle06}. Depolarization could originate due to a combination of ---
frequency independent beam depolarization arising from turbulent magnetic
fields on scales smaller than the telescope beam, and frequency-dependent
Faraday depolarization due to the nature of the Galactic magneto-ionic medium.
The latter can also give rise to strong spatial variations of the polarized
synchrotron emission making it difficult to constrain its contribution at CMB
frequencies.  As a result, typically, the region $|b| \lesssim 20^\circ$
(roughly 30--40\% of the total sky-area) is masked from CMB analyses. Among
other things, this results in increased cosmic variance in the measurement of
CMB angular power spectra on large angular scales which is of interest for
next-generation CMB space missions.

The only all-sky Galactic Faraday depth (FD) distribution is available via
interpolation of spatially discrete FDs measured towards extragalactic sources
using Bayesian analysis \citep{opper12}. The Galactic FD varies widely from
$-500$ to $+500\,\rm rad\,m^{-2}$ in the region $|b| \lesssim 30^\circ$. In
regions of such high FD, frequency-dependent Faraday depolarization increases
towards lower frequencies and therefore the polarized Galactic emission in
previous polarization surveys near 1.4~GHz were found to be depolarized. To
mitigate the effects of Faraday depolarization and to accurately constrain the
Galactic linearly polarized synchrotron emission at CMB frequencies, only
recently efforts have turned towards the frequency range 2 to 10~GHz and to
using wider frequency coverages. The frequency range 2.176 to 2.4~GHz is
covered by the S-PASS and the 4.5 to 5.5~GHz is covered by the C-BASS. The
advantage of performing polarization surveys at frequencies higher than 2~GHz
to reduce the effects of Faraday depolarization has been demonstrated by 
the S-PASS data \citep{carre13, krach18}.

However, for $|b| \lesssim 30^\circ$, strong Faraday depolarization means that
the polarized structures observed using the relatively narrow band of S-PASS
are also expected to show strong spectral variations. A large frequency gap
between S-PASS and C-BASS would make combining the S-PASS and C-BASS data a
challenging proposition. In order to robustly recover the polarized emission at
low Galactic latitudes, wide bandwidth polarization survey using the \mmska,
covering the 1.7 to 3.5~GHz range, will be crucial. In fact, the wide frequency
coverage also allows us to constrain the Galactic total synchrotron emission
spectrum using physically motivated models all over the Southern sky.

\section{Synchrotron and free--free foregrounds} \label{sec:radio_fg}

Each of the foreground emission components have distinct spectrum and dominates
at different frequency regimes. Synchrotron emission typically has a steep
power-law spectrum and dominates at frequencies $\lesssim10$ GHz, while the
thermal free--free emission has a flatter power-law spectrum which becomes
significant above $\sim$5 GHz. The AME spectrum has varying peaks in the
frequency range $10 \text{--} 50$ GHz. Thermal dust emission follows a modified
Planck spectrum which starts to dominate above $\sim$100 GHz. We refer
interested readers to \citet{thorn17} and \citet{jones18} for details on
foreground emission components. At frequencies above 10 GHz, where a wealth of
foreground components become important, their modeling increasingly suffers
from degeneracies giving rise to large uncertainties \citep{planck2015X}. Thus,
to independently constrain the synchrotron and free--free emission components,
observations below $\sim$10 GHz are necessary, while AME and thermal dust
emission are better constrained at frequencies above 10 and 100~GHz,
respectively.

In order to constrain various low-frequency foreground components, dedicated
surveys to map both total and polarized emissions are being performed, e.g.,
the C-BASS and the S-PASS at frequencies below 10~GHz, and the Q-U-I JOint
TEnerife \citep[QUIJOTE;][]{quijote15a, quijote15b} and
GreenPol\footnote{https://www.deepspace.ucsb.edu/projects/greenpol} experiments
in the 10 to 40 GHz frequency range. Short descriptions of these recent
efforts are presented in Appendix~\ref{sec:survey}.  In this paper, we will
focus on constraining the total radio continuum emission below
$\lesssim10$\,GHz and the polarized synchrotron emission from our Galaxy where
observations in the S-Band with the \mmska\,will play an important role.

\subsection{Total intensity spectrum} \label{sec:synchrotron}

In the interstellar medium (ISM), the synchrotron-emitting relativistic
electrons, also referred to as cosmic ray electrons (CREs), are believed to be
accelerated in the shock fronts of supernova explosions via diffusive shock
acceleration and are injected into the ISM with a power-law energy spectrum
\citep{bell04, jones11, edmon11, kang12}. As a consequence, the synchrotron
emission arising from freshly injected CREs has a power-law frequency spectrum
with slope $\bnt$.\footnote{In this paper, we will represent the frequency
spectrum-slope ($\beta$) in terms of the brightness temperature ($T_{\rm B}$),
i.e., $T_{\rm B}(\nu) \propto \nu^\beta$.  The corresponding flux density
spectrum $S(\nu)$ is given by, $S(\nu) \propto \nu^{\alpha}$ and $\alpha$ is
related to $\beta$ as $\alpha = \beta + 2$.} The value of $\bnt$ depends on the
Mach number ($M$) of the shock as, $\bnt = -(5M^2-1)/(2M^2 -2)$
\citep{bland87}. For strong shocks, i.e., $M\gg1$, $\bnt \approx -2.5$.
However, since the Mach number depends on the sound speed and thereby the gas
density of the medium where CREs are injected, $\bnt$ could lie in the 
range $-2.5$ to $-3$ \citep[see, e.g.,][]{capri14, park15}.

Since injection, the CREs are subjected to energy dependent losses via various
mechanisms, such as, ionization, relativistic bremsstrahlung, synchrotron and
inverse-Compton \citep[see e.g.,][]{longa11}. These losses affect different
parts of the frequency spectrum and lead to the injected power-law synchrotron
spectrum to be modified smoothly over a frequency range.  Synchrotron and
inverse-Compton losses typically dominate at frequencies $\gtrsim 2$ GHz which
steepens the spectrum, and the nature of steepening depends on the process of
particle injection. For example, beyond a break frequency $\nu_{\rm br}$ the
spectrum steepens from $\bnt$ to $\bnt - 0.5$ in the case of continuous
injection of CREs in an synchrotron emitting volume \citep{pacho70}. In the
scenario when CREs in a volume are injected at a single epoch, the synchrotron
spectrum either falls off as a power-law with index $(4\,\bnt -1)/3$ for $\nu >
\nu_{\rm br}$ \citep[KP model;][]{karda62, pacho70} or develops an exponential
cut-off at a cut-off frequency $\nu_{\rm c}$ \citep[JP model;][]{jaffe73}.

Thus, to describe physical models of CRE energy losses, the synchrotron
brightness temperature ($T_{\rm B, syn}$) spectrum can be represented as,
\begin{equation} 
T_{\rm B,syn}(\nu) = T_{\rm syn,0}\,\frac{(\nu/\nu_0)^{\bnt}}{\left[1 +
(\nu/\nu_{\rm br})^\gamma\right]}\,{\rm e}^{-(\nu/\nu_{\rm c})}.
\label{eq:synchrotron} 
\end{equation} 
Here, $T_{\rm syn,0}$ is the brightness temperature normalization of the
synchrotron emission at a pivot frequency $\nu_0$ chosen such that $\nu_0 <
\nu_{\rm br}$ and $\nu_0 < \nu_{\rm c}$, and $\gamma$ is the spectral curvature
parameter. Note that, in the case when $\nu_{\rm c} \rightarrow \infty$,
$\gamma = 0.5$ represents the scenario of continuous injection of CREs and
$\gamma = (-\bnt + 1)/3$ represents the KP model. The JP model of CRE
injections is given by a finite value of $\nu_{\rm c} \lesssim \nu_{\rm br}$. 

The break- and cutoff-frequencies originating due to synchrotron and/or
inverse-Compton cooling depends on the total magnetic field strength, the
interstellar radiation field and age of the CREs in a way that, the older the
CREs, the lower are $\nu_{\rm br}$ and $\nu_{\rm c}$. The bulk of the CREs in
galaxies is produced in galactic discs, roughly in a $\pm 500$ pc region around
galactic mid-plane. The CREs then propagate away from the galactic disc into
the galactic halo via various transport mechanisms, such as, diffusion,
streaming at Alfv\'en speed, galactic winds and/or advection \citep[see
e.g.,][]{heese16, heese18}. Depending on the dominant transport and diffusion
mechanisms, CREs can take $\sim10^6 \text{--} 10^8$ years to reach up to
$\sim2$ kpc or more above the mid-plane \citep{kraus18}. Such transport
time-scale of CREs in Galactic magnetic field strengths of $5 \text{--} 15
\,\umu$G would result in $\nu_{\rm br}$ and $\nu_{\rm c}$ to typically lie in
the range 2--10 GHz at high Galactic latitudes, $|b|\gtrsim30^\circ$. At lower
$|b|$, the CREs are expected to be young, and therefore $\nu_{\rm br}$ and
$\nu_{\rm c}$ can lie at frequencies above 10 GHz. At an average, we expect
break frequencies to decrease with increasing Galactic latitude $|b|$.

Spatially varying ISM conditions and magnetic field strengths give rise to
locally varying breaks in the synchrotron spectrum which when averaged over
large volume by a telescope beam could smooth-out the synchrotron spectrum
\citep{basu15a}. Also, mixing of different CRE populations via propagation and
steady injection in the ISM would further smooth out any sharp spectral
fall-off of the synchrotron emission as predicted by the KP and the JP models.
Hence, in the Galactic ISM, we expect $\gamma$ to lie in the practical range 0
and $+2$. Because of increasing CRE age and decreasing CRE injection rates with
increasing $|b|$, steeper break in the synchrotron emission spectrum at high
Galactic latitudes is expected. That means, in our parametrization for the
synchrotron emission, $\gamma$ is expected to increase with Galactic latitude
$|b|$.

The Galactic free--free emission at frequencies above $\sim$1~GHz is expected
to be optically thin and follows a power-law spectrum with a spatially constant
free--free spectral index $\beta_{\rm ff}\approx$$-2.1$. Strictly speaking,
$\beta_{\rm ff}$ is expected to decrease slightly with frequency due to
frequency and electron temperature ($T_{\rm e}$) dependent Gaunt factor
\citep{benne92, dicki03}. At the frequencies of our interest, and for the
typical range of $T_{\rm e}$ between 4000 and 10\,000 K in the Galactic ISM,
the widely used approximate value of $\beta_{\rm ff} = -2.1$ leads to 
systematic underestimation of the free--free emission by $\lesssim 2\%$
\citep{dicki03}. Therefore, below 10~GHz, the total Galactic radio continuum
emission {$T_{\rm B}(\nu)$} can be well described by,
\begin{equation} 
\begin{split}\label{eq:model} 
T_{\rm B}(\nu) & = T_{\rm B, syn}(\nu) + T_{\rm B, ff}(\nu) \\ & = T_{\rm
syn,0}\,\frac{(\nu/\nu_0)^{\bnt}}{\left[1 + (\nu/\nu_{\rm
br})^\gamma\right]}\,{\rm e}^{-(\nu/\nu_{\rm c})} + T_{\rm
ff,0}\,(\nu/\nu_0)^{-2.1}, 
\end{split} 
\end{equation} 
with six free parameters $T_{\rm syn, 0}$, $T_{\rm ff,0}$ (normalization of the
free--free emission at $\nu_0$), $\bnt$, $\gamma$, $\nu_{\text{br}}$ and
$\nu_{\rm c}$. $T_{\rm B,ff}$ is the brightness temperature of the free--free
emission. Large frequency coverage thereby provides the opportunity of
constraining the Galactic synchrotron and free--free emission components
directly from the total intensity radio continuum spectrum.

\subsection{Complexities in the synchrotron polarization spectrum} \label{sec:depol}

Since the total intensity spectra of the different foreground components are
smooth, once they are constrained at lower radio frequencies, e.g., between 5
and 40 GHz, it is simple to extrapolate them to higher frequencies around the
{dominant }CMB emission, i.e., near 100 GHz. Extrapolation of the
polarized emission, especially the polarized synchrotron emission measured at
frequencies below 10 GHz is non-trivial due to the effects Faraday rotation and
Faraday depolarization have on the polarized emission spectrum. 

Faraday rotation describes the effect wherein the plane of polarization of a
linearly polarized signal is rotated {by an angle $\Delta \theta$} while
traversing in a magneto-ionic medium and the observed angle of polarization
($\theta$) at a wavelength $\lambda$ is given by,
\begin{equation} 
\theta(\lambda) = \theta_0 + \Delta \theta = \theta_0 + {\rm FD}\,\lambda^2.  
\end{equation} 
Here, $\rm FD$ is the Faraday depth and $\theta_0$ is the intrinsic
polarization angle of {the linearly polarized signal, typically
originating from synchrotron emission.} In a purely Faraday rotating medium,
the observed polarization angle varies linearly with $\lambda^2$, and $\rm FD$
is given by its slope, which is also referred to as the Faraday rotation
measure and is commonly denoted as $\rm RM$. $\rm FD$ is the integral of the
magnetic field component along the line of sight ($B_\|$) weighted by the
thermal electron density ($n_{\rm e}$), from the polarized source to an
observer, and is given by, 
\begin{equation} 
{\rm FD} = 0.812\,\int\limits_{\rm source}^{\rm observer}\,\left(\dfrac{n_{\rm
e}(l)}{\rm cm^{-3}} \right)\,\left(\dfrac{B_\|(l)}{\rm \umu
G}\right)\,\left(\dfrac{dl}{\rm pc}\right)\,\, {\rm rad\,m^{-2}}.  
\label{eq:fd} 
\end{equation} 
In such a situation, the complex fractional polarization\footnote{Fractional
polarization is defined as $p = PI/I_{\rm syn}$, where $PI$ and $I_{\rm syn}$
are the observed linearly polarized and total synchrotron intensities,
respectively.} ($p$) varies with $\lambda$ as, 
\begin{equation}
p(\lambda) = p_0\,{\rm e}^{2\,i\,(\theta_0\,+\,{\rm FD}\,\lambda^2)}.  
\end{equation}

In typical astrophysical plasmas, e.g., those encountered in the Galactic ISM,
Faraday rotating and synchrotron emitting plasmas are mixed, often leading to
breakdown of the linear $\theta$ vs. $\lambda^2$ relation.  {In such
scenarios, a part of the integral for FD in Eq.~\eqref{eq:fd} is also
contributed by a synchrotron emitting medium. The FD contributed by the part of
a line of sight which is simultaneously Faraday rotating and synchrotron
emitting will be denoted as $\rm FD_{em}$. That is, the limits of the line
integral in Eq.~\eqref{eq:fd} spans from far to the near side of the
synchrotron emitting volume only. In this paper, FD always represents the
Faraday depth along the entire line of sight as given by Eq.~\eqref{eq:fd},
unless specified otherwise.}

Further, depending on how the magnetic fields and free electrons are
distributed in the plasma, $p$ varies non-linearly with $\lambda^2$ and its
analytical solutions under certain approximations are derived in detail by
\citet{sokol98}. The variation of $p$ as a function of $\lambda$ in a volume
where synchrotron emission and Faraday rotation originate in uniform magnetic
fields, known as the `Burn slab' model \citep{burn66, sokol98}, is given by,
\begin{equation} 
p(\lambda) = p_0\,\dfrac{\sin\, {\rm FD_{em}}\,\lambda^2}{{\rm FD_{em}}\,\lambda^2}\,
{\rm e}^{2\,i\,\left(\theta_0\,+\,\frac{1}{2}\,{\rm FD_{em}}\,\lambda^2\right)}.
\label{eq:burn} 
\end{equation}

If the volume is also clumpy and contains both turbulent and regular magnetic
fields within which $\rm FD$ has a dispersion $\sigma_{\rm FD}$, known as the
internal Faraday dispersion model, the variation of $p$ with $\lambda$ is given
by \citep{sokol98}, 
\begin{equation} 
p(\lambda) = p_0\,{\rm e}^{2\,i\,\theta_0}\,\left(\dfrac{1 \, - \,
\exp[-(2\,\sigma_{\rm FD}^2\,\lambda^4 - 2\,i\,{\rm FD_{em}}\,\lambda^2)]}
{2\,\sigma_{\rm FD}^2\,\lambda^4 - 2\,i\,{\rm FD_{em}}\,\lambda^2} \right).  
\label{eq:int_disp} 
\end{equation}

{Note that, it is possible that the magneto-ionic media described by
equations~\eqref{eq:burn} and \eqref{eq:int_disp} lies behind purely Faraday
rotating media with Faraday depth $\rm FD_{fg}$. Such cases be accounted
for by multiplying these two equations by $\exp({2\,i\,{\rm FD_{fg}} \,
\lambda^2})$. FD defined in Eq.~\eqref{eq:fd} is then $\rm FD = FD_{fg} +
FD_{em}$.}

If a foreground Faraday rotating medium only contains turbulent magnetic
fields, known as the external Faraday dispersion model, $p$ varies with
$\lambda$ as \citep{sokol98}, 
\begin{equation} 
p(\lambda) = p_0\,{\rm e}^{-2\,\sigma_{\rm FD}^2\,\lambda^4}\,{\rm e}^{2\,i\,
(\theta_0\,+\,{\rm FD}\,\lambda^2)}.  
\label{eq:ext_disp}
\end{equation}

The dispersion of FD, $\sigma_{\rm FD}$, within the 3-dimensional volume probed
by a telescope beam is related to the turbulent magnetic field along the line
of sight, where $B_\|$ has root-mean square (rms) $b_\|$, and properties of the
magneto-ionic medium as, 
\begin{equation} 
\sigma_{\rm FD} = 0.812\,\left(\dfrac{\langle n_{\rm e}\rangle}{\rm
cm^{-3}}\right)\, \left(\dfrac{b_\|}{\rm \umu G}\right)\,\left(\dfrac{{f_{\rm
V}^{-1}}\, L \,l_{\rm c}}{\rm pc^2}\right)^{1/2}\,\,{\rm rad\,m^{-2}}.  
\end{equation} 
Here, $\langle n_{\rm e}\rangle$ is the average density of thermal electrons
with a volume filling factor of $f_{\rm V}$. $L$ is the path length through the
ionized medium and $l_{\rm c}$ is the correlation length of the product $n_{\rm
e}\,b_\|$, referred to as the turbulent cell size. Depending on how turbulent
energy is injected into the ISM of galaxies, $l_{\rm c}$ can be a few pc when
driven by protostellar outflows, 50--100 pc when driven by supernovae, or a few
kpc when driven by Galactic rotation, superbubble and/or Parker instability
\citep[see, e.g.,][]{armst95, elmeg04, haver08, krumh18}.  Therefore, for
typical $b_\| \approx 10~\umu$G and $L \approx 5$~kpc, and depending on the
location within the Galaxy, $\sigma_{\rm FD}$ can range from up to
$\sim$10~$\rm rad\,m^{-2}$ at high Galactic latitudes to 50--200~$\rm
rad\,m^{-2}$ at low Galactic latitudes.

\citet{sulli17} suggests a general description of all the above depolarizations
in the presence of both regular and turbulent magnetic fields as,
\begin{equation} 
p(\lambda) = p_0\,\dfrac{\sin\, {\rm \Delta FD}\,\lambda^2}{{\rm \Delta
FD}\,\lambda^2}\, {\rm e}^{-2\,\sigma_{\rm FD}^2\,\lambda^4}\,{\rm
e}^{2\,i\,(\theta_0\,+\,{\rm FD}\, \lambda^2)}.
\label{eq:gen_depol} 
\end{equation} 
Here, $\Delta {\rm FD}$ is the gradient of Faraday depth caused by variation of
regular magnetic fields within the telescope beam.

In the case when the telescope beam probes a large volume of an astrophysical
system, a combination of the above mentioned plasma conditions are observed.
Therefore, in order to model the observed variation of $p(\lambda)$, one
linearly combines these models and direct fitting of the Stokes $Q$ and $U$
intensity spectra is performed known as the Stokes $Q$, $U$ fitting. Of late,
with the acquisition of broad band radio polarimetric data, the technique of
Stokes $Q$, $U$ fitting has been applied to infer properties of the
magneto-ionic medium in different types of extragalactic objects, e.g., active
galactic nuclei \citep{sulli12, sulli17, paset18}, external galaxies
\citep{shnei14, mao15} and high redshift galaxies \citep{kim16, mao17}.

\begin{table*} 
\caption{Expected specifications of the \mmska.} \begin{center}
\begin{tabular}{@{}ll@{}} 
\hline 
Location & Karoo, South Africa\\
Coordinates & $30^\circ\,43^\prime\,4.69^{\prime\prime}$S, $21^\circ\,24^\prime\,47.02^{\prime\prime}$E\\ 
Antenna & 15-m offset-Gregorian \\ & \\ 
\end{tabular} \\

\begin{tabular}{@{}lcc@{}} 
\hline
& S-Band & Ku-Band \\ 
\cline{2-3}\\ 
Rx frequency & 1.7--3.5 GHz & 12--18 GHz \\ 
Digitizer bandwidth per Stokes & 1.75 GHz (12-bit) & 3 GHz (12-bit) \\ 
Frequency channels & 1024 or 2048 & 2048 \\ 
Polarizer & Linear & Circular \\ 
\multicolumn{1}{}{} &
	\multicolumn{2}{c}{~~{($<-35$ dB cross coupling before calibration)}}\\ 
\multicolumn{1}{l}{Stokes} &
\multicolumn{2}{c}{~~~$I, Q, U, V$}\\ 
\\ 
	         & $T_{\rm sys}<20$ K & $T_{\rm rec}<18$ K \\ 
Beam FWHM$^\dagger$ & 50--25 arcmin & 7--4.6 arcmin \\ 
{Total intensity confusion}$^\dagger$ & 300--50 mJy (16--2 mK) &  1.7--0.6 mJy (80--28 $\umu$K)$^a$\\ 
{Stokes $Q$ \& $U$ confusion}: &   &  \\ 
~~~~~Polarized sources$^{\dagger,b}$ & 1.7--0.23 mJy (90--9 $\umu$K)  & 7.5--2.4 $\umu$Jy (0.40--0.12 $\umu$K) \\ 
~~~~~Polarization leakage$^{\dagger,c}$ & 0.6--0.1 mJy (32--4 $\umu$K) & 3.4--1.2 $\umu$Jy (0.16--0.06 $\umu$K) \\ 
Typical integration required:$^\ddagger$ & & \\
~~~~~~Total intensity & 30 second & $\approx10$ minute \\ 
~~~~~~Stokes $Q$ \& $U$ & 2 hour & -- \\ 
\\ 
Faraday depth FWHM & $\approx 140$ rad\,m$^{-2}$ & $\approx10\,000$ rad\,m$^{-2}$ \\ 
Maximum Faraday depth & $\approx \pm4\times10^4$ rad\,m$^{-2}$ & $\approx \pm 6.5\times10^5$ rad\,m$^{-2}$\\ 
Faraday depth scale & $\approx 450$ rad\,m$^{-2}$ & $\approx 10\,000$ rad\,m$^{-2}$\\ 
\hline 
\end{tabular} 
\end{center}
\begin{flushleft} 
$^\dagger$ The range corresponds to values at the lowest and the highest
frequency end of the receiver.\\ 
$^a$ The confusion noise limit at Ku-Band was estimated by assuming the same source
distribution as that at S-Band and a typical spectral index of $-2.7$ for
extragalactic sources.\\
$^b$ Confusion from polarized sources is calculated based on \citet{loi19}.\\
$^c$ Assuming a conservative {on-axis instrumental leakage} up to 0.2\% on unpolarized regions.\\ 
$^\ddagger$ For reaching confusion noise over 10~MHz channel-width per beam at S-Band
and 1~GHz channel-width at Ku-Band.  
\end{flushleft} 
\label{tab:specs} 
\end{table*}

The technique of Stokes $Q$, $U$ fitting is a powerful tool for extrapolating
polarized emission observed at low radio frequencies to arbitrarily high
frequencies. However, for robust fits, large $\lambda^2$-coverage through broad
bandwidth polarization measurements is necessary, without which the fitted
models and their parameters could be degenerate giving rise to large systematic
uncertainties. Further, since Faraday depolarization depends strongly on
$\lambda$, polarization measurements at small $\lambda$ or high frequencies
($\gtrsim5$ GHz) often do not adequately capture the depolarization features in
the polarized emission spectrum introduced by multiple components and are
therefore insufficient. At low Galactic latitudes, $|b|\lesssim 30^\circ$,
strong Faraday depolarization gives rise to strong spectral and spatial
variations of the polarized synchrotron emission. {Therefore,} at
longer $\lambda$, i.e., frequencies $\lesssim1$ GHz, Faraday depolarization is
severe leading to either very little polarized signal or the observed polarized
emission originating locally \citep[e.g.,][]{jelic15}. These issues
necessitates broadband polarization survey in the frequency range 2 to 5~GHz.

\section{The SKA-MPG Telescope} \label{sec:m2skad}

\begin{figure} 
\begin{centering} 
\begin{tabular}{c}
{\mbox{\includegraphics[width=9.0cm, trim=1.6cm 1.2cm 1.2cm 1cm, clip]{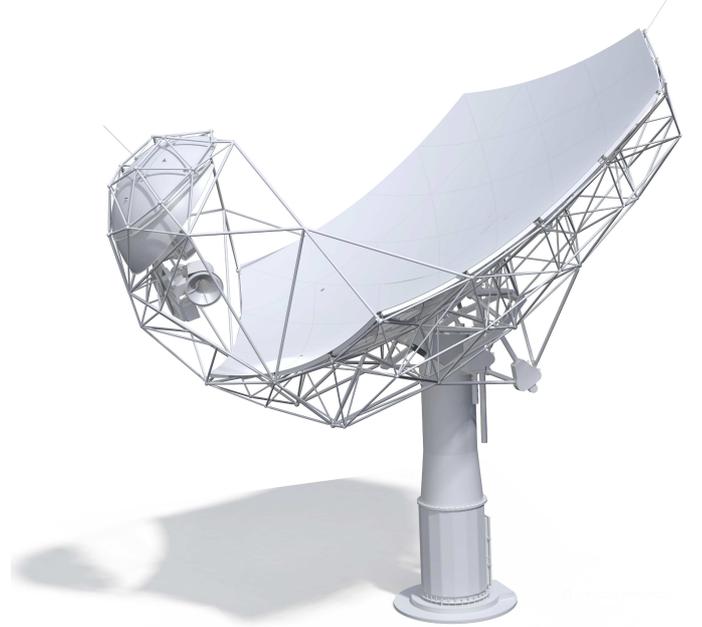}}}\\
\end{tabular} 
\end{centering} 
\caption{A schematic view of the \mmska\ which is being constructed in the
Karoo desert, South Africa.}
\label{fig:dish} 
\end{figure}

The \mmska, shown in Fig.~\ref{fig:dish}, is a \mbox{15-m} aperture
offset-Gregorian telescope being constructed by the MT Mechatronics GmbH in
collaboration with the Max-Planck-Institut f\"ur Radioastronomie (MPIfR) of the
Max Planck Society (MPG). This telescope is the first prototype dish of the
South African Square Kilometre Array (SKA)-MID component that is being
assembled in the Karoo desert, South Africa. The location of the telescope is
offset with respect to the positions of the MeerKAT- and SKA-MID-configuration
and is specially suited for a critical system design review and an independent
scientific programme. The telescope will be equipped with S- and Ku-Band
receiver systems, both developed by the MPIfR (for details see
Table~\ref{tab:specs}). The S-Band receiver covers 1.7 to 3.5~GHz and the
Ku-Band receiver covers 12 to 18~GHz, and depending on the scientific merit
additional frontends could be installed in the future.

In general, the benefits of an offset-Gregorian design are: an increased gain
due to no aperture blockage, a reduction of standing waves in the aperture, and
a reduction of stray radiation and possibly radio frequency interference (RFI).
A disadvantage of this design is an asymmetric primary beam that may require a
more complex calibration procedure. However in the telescope design phase
special care has been taken to optimize the primary beam properties.
Theoretical modelling indicates a well behaved primary beam having an
ellipticity of 0.022, and the first and the second sidelobes are at $-24$ and
$-32.8$ dB, respectively, at the centre frequency 2.6~GHz in S-Band. Until a
full characterization of the system, only the theoretical limits are provided.
The collecting area of the \mmska\,depends on the optical path between the
primary and the secondary reflector system and its shaped optics is optimized
to an effective area equivalent to a 15-m dish. This will provide us angular
resolutions between 50 to 25 arcmin at the lowest and the highest frequency of
the S-Band, respectively (see Table~\ref{tab:specs}). Therefore, the lowest
frequency limits the angular resolution of sky survey using the \mmska\ to 1
degree. 

Both {S- and Ku-Band} receiver systems follow the classical design of
cryogenic cooled receivers, with a room-temperature linear feed for the S-Band
and a cryogenic circular feed for the Ku-Band system. Both systems are equipped
with a 12-bit, temperature stabilized, internal digitizer that operates on $2
\textrm{ Stokes} \times 1.75$~GHz and $2 \textrm{ Stokes} \times 3$~GHz
bandwidth in S- and Ku-Band, respectively. Thus, the \mmska\ will provide an
instantaneous bandwidth of 1.75~GHz in the S-Band and 3~GHz in the Ku-Band.
After the digitizer, the data stream is structured by the packetizer, with an
optional bandwidth reduction, before it is transported to the central
processing building. In the final post processing step, the properties of the
data products are freely choosable within computational limits and can be
optimized for each individual observation. As an example, the variable
configuration allows for a maximum of 2048 channels over the entire frequency
bands. The anticipated system stability may reach a dynamic range of the order
of 35 dB or better and a system temperature ($T_{\rm sys}$) below 20\,K in
S-Band {and receiver temperature ($T_{\rm rec}$) below 18\,K in
Ku-Band}. The expected setup of the data products are summarized in
Table~\ref{tab:specs}. 

\begin{figure} 
\begin{centering}
\begin{tabular}{c}
{\mbox{\includegraphics[width=8.5cm]{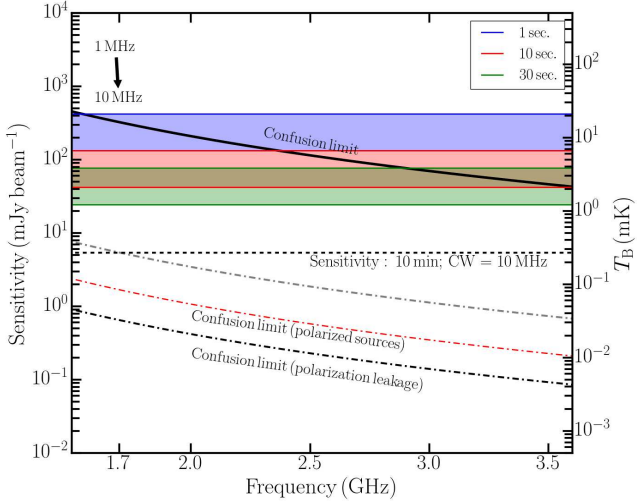}}}\\
\end{tabular} 
\end{centering} 
\caption{Expected sensitivity limits of \mmska. The solid black curve show the
total intensity confusion limit and the black dash-dot curve is the confusion
in Stokes $Q$ and $U$ assuming 0.2\% instrumental polarization leakage.  The
red dash-dot curve shows the confusion arising from polarized sources. The grey
dash-dot curve is for $3\times$ the total confusion noise in polarization. The
various bands show the range of sensitivity expected for integration time of 1
second (blue), 10 seconds (red) and 30 seconds (green).  The upper edge for
each band is averaging over 1 MHz and the bottom is for averaging over 10 MHz.
The black dashed line is the sensitivity for 10 MHz averaging and an
integration time of 10 minutes {(see Section~\ref{sec:sensitivity})}.}
\label{fig:sensitivity}
\end{figure}

\subsection{Sensitivity} \label{sec:sensitivity}

In Fig.~\ref{fig:sensitivity}, we show the expected sensitivity of the \mmska\,
over its S-Band frequency coverage for $T_{\rm sys} = 20$ K. The different
coloured bands are for different integration times per resolution element
averaged over 1 MHz (top edge of each colour) to 10 MHz (bottom edge of each
colour). Low-frequency single-dish surveys are limited by confusion noise in
total intensity arising from unresolved sources and is expected to be in the
range of 2 to 15 mK (the solid black line in Fig.~\ref{fig:sensitivity}) at the
frequency and resolution of the \mmska. The confusion limit can be achieved in
just 30 seconds for 10-MHz wide frequency channels over the entire S-Band.
Assuming a close to uniform sky-coverage, a total-intensity-confusion limited
survey of the Southern sky (an area of $\approx21\,000\,\deg^2$), can be
performed in about 300 hours\footnote{{Here we have considered the
possibility of a non-uniform sky coverage in terms of the integration time due
to scanning overlaps. The estimated time also includes, in about 15\% of the
sky, at an average, the integration time is increased by a factor of five.}}
(including astronomical calibration overhead of 20\%) making \mmska\ an
excellent, fast, sky-survey instrument. In polarized intensity, however, due to
the lack of bright polarized sources, the confusion noise is expected to be at
$<0.1$ mK level \citep[following][]{stil14, loi19} over the entire S-Band and
hence deeper sensitivity to polarized emission can be achieved with longer
integration.

{In S-Band, a sensitivity of} $\sim$2 mK is sufficient for detecting
diffuse polarized emission from the entire Southern-sky as suggested by the
S-PASS at 2.3 GHz \citep[see][]{krach18, carre19}. However, to robustly apply
the Stokes $Q,U$ fitting technique, a signal-to-noise ratio (SNR) $>5$ is
required per frequency channel \citep{schni18}. Hence, the target is to achieve
sensitivity of 0.3 mK in Stokes $Q$ and $U$, averaged over 10 MHz channel-width
across the entire S-Band. We note that, due to frequency dependent variation of
the polarized intensity at $|b| \lesssim 30^\circ$, SNR per 10-MHz channel
could vary. At  $|b| \gtrsim 30^\circ$, as Faraday depolarization is expected
to be low, the polarized intensity is expected to follow a close to power-law
spectrum. Therefore, the Galactic polarized emission in the higher
frequency-end of S-Band could be weaker compared to the representative value of
sensitivity at 2.3~GHz used by us. In those regions, to achieve higher SNR, we
can afford to average to 40-MHz channel-widths or more without being
significantly affected by channel-width depolarization.

Achieving a sensitivity of 0.3 mK in the polarized intensity, also necessary
for better foreground removal for CMB related studies, will require about 10
minutes averaged over 10 MHz (see Fig.~\ref{fig:sensitivity}) per resolution
element. A close-to 10 MHz channel-width will be achieved by averaging 10--12
channels, i.e., about 200 frequency channels {spread} over the S-Band
coverage of the \mmska. To reach the requisite sensitivity, the Southern sky
will be covered 20 times, each time performing a 300-hour
total-intensity-confusion limited survey.

\begin{figure*} 
\centering
\begin{tabular}{cc}
{\mbox{\includegraphics[width=8cm]{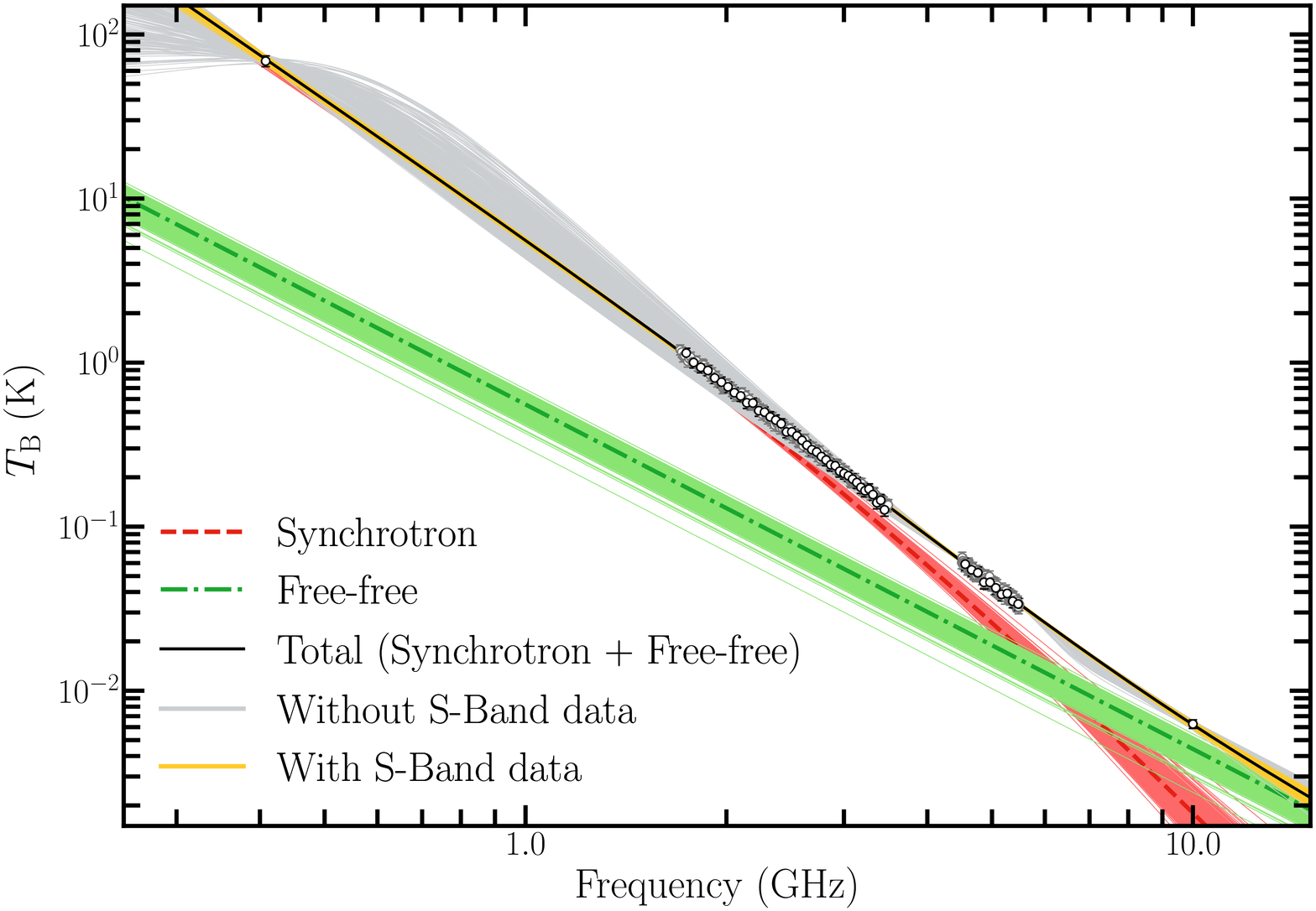}}} &
{\mbox{\includegraphics[width=8cm]{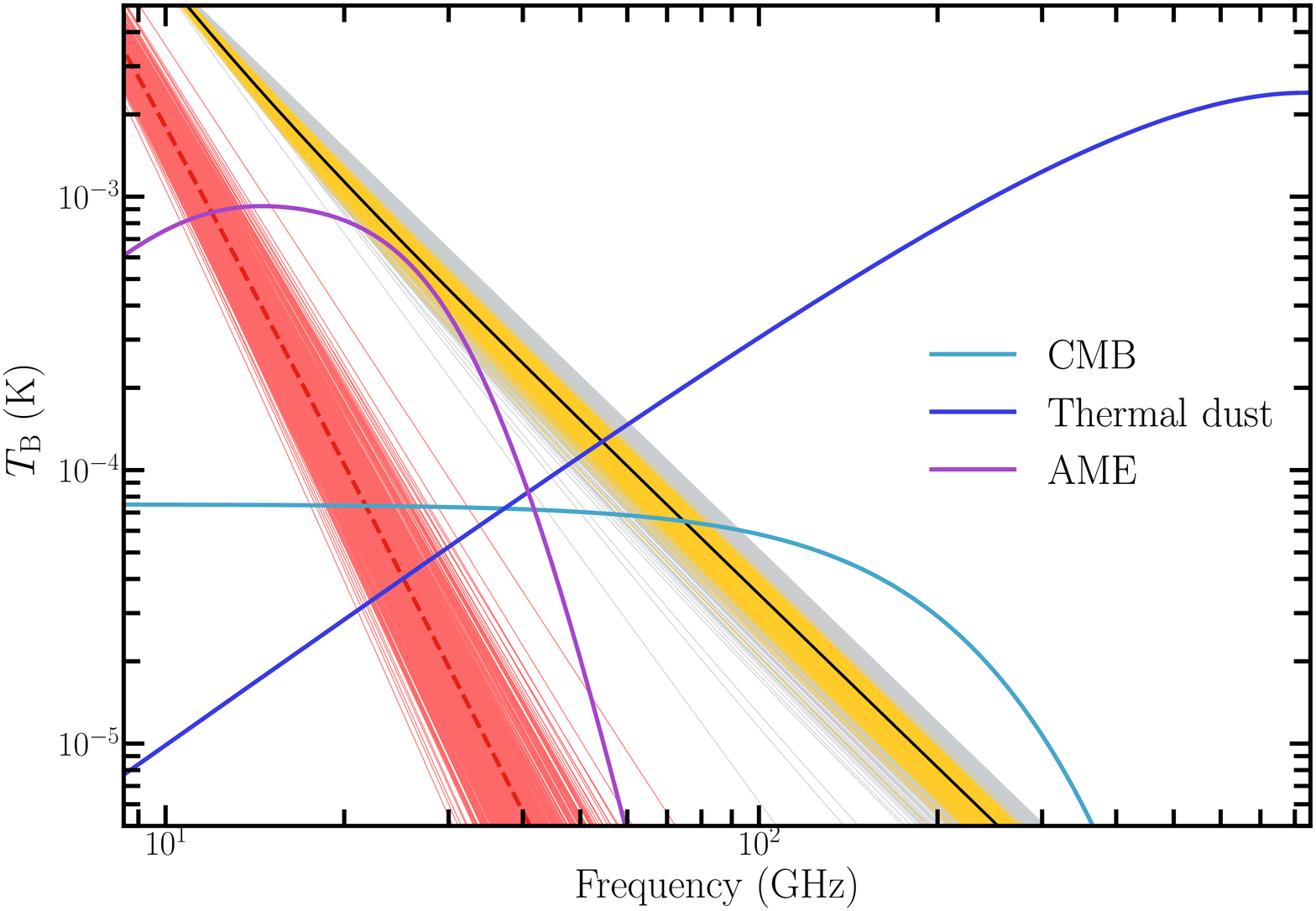}}}\\ 
\end{tabular}
\caption{{\it Left}: An example total radio continuum frequency spectrum is
shown as the solid black line. The red dashed line is the synchrotron emission
spectrum with $\nu_{\rm br} = 4$ GHz normalized to 5 K at 1 GHz, the green
dot-dashed line is the free--free emission spectrum with a thermal fraction of
10\% at 1 GHz. The data points are simulated measurements at 408\,MHz, S-Band,
C-Band, and 10\,GHz with noise. The coloured lines show fits to 500 random
realizations of simulated data using Eq.~\eqref{eq:model}. The yellow and grey
lines are for fits with and without the S-Band data, respectively.  {\it
Right}: Fits in the left panel extrapolated to higher frequencies. To highlight
the recovered synchrotron + free--free emission, the extrapolated free--free
component fits are not shown. For comparison, we show the spectrum of the CMB
total intensity rms as the light-blue line. The foreground AME and thermal dust
emission is shown as the magenta and dark blue lines, respectively. All the
other lines have the same meaning as in the left panel.} 
\label{fig:spectra} 
\end{figure*}

\subsection{Observational challenges} \label{sec:obs}

Single dish telescopes are ideal for performing large sky-area surveys as they
are sensitive to diffuse emission on all angular scales, unlike interferometers
which suffer from missing emission structures on scales larger than the
resolution of the shortest baseline. However, single dish observations face
challenges from systematics arising because of the additive nature of recording
signals. A full description of the survey strategy and characterization of the
telescope properties are beyond the scope of this paper. They will be presented
in detail in another paper after the construction of \mmska\ is completed and
is ready for scientific commissioning. Here we discuss in brief about the
features of the \mmska\ in coping with the observational challenges.

{\it Scanning:} In order to recover diffuse emissions and map them accurately,
usual scanning artefacts are mitigated by performing cross-linked scans and
applying the technique of basket-weaving. To optimize time efficiency and
telescope slewing for large sky-area surveys, long scans along the telescope's
azimuthal direction at a fixed elevation is performed \citep[see,
e.g.,][]{carre19, jones18}. This type of long azimuthal scans typically results
in non-uniform sky coverage {in terms of integration time}. One way
to achieve close to uniform sky coverage is to perform azimuthal scans fixed at
several elevation angles and driving the telescope at faster speeds in regions
of multiple scan overlap. The \mmska\ offers a maximum slewing speed of
$3^\circ \rm \,s^{-1}$ along the azimuthal axis and $1^\circ \rm \, s^{-1}$
along the elevation axis. The automatic control unit on the telescope allows
position determination on a $50\,\umu$s time-scale, allowing us to achieve
sky-scanning speed close to the maximum slewing speed. However, practical
time-scales for telescope position determination depend on the dish tracking
operations and the frequency of the status message update. These can be
optimized based on the needs of the observing programme. We are currently
exploring different scanning strategies to optimize the total survey time to
cover the sky as uniformly as possible.  Depending on the scanning strategy, it
is possible that the time-scale of each total-intensity-confusion limited
survey presented in Section~\ref{sec:sensitivity} can go up by up to a factor
of two.

{\it Ground spillover:} Single dish telescopes are prone to pick up stray
emission from the ground, especially when observing at low telescope elevation.
This is even more true for an offset-Gregorian feed as used for the \mmska. The
ground spillover can be significantly reduced by extending the sub-reflector to
the bottom. For the \mmska\ an extension of $40^\circ$ has been selected
(I.~P.~Theron, SKA-TEL-DSH-0000018, Rev 2, Dish Optics Selection Report) which
will reflect any spillover towards the sky. We therefore believe that the pick
up emission from the ground at low telescope elevations will be low for the
survey.

{\it Calibration:} Novel designing of the S-Band receiver on the \mmska\
provides spectroscopic Allan time of $\gtrsim1000$ seconds having wide band
gain stability with rms better than 0.03\%. This considerably reduces the
frequency of performing calibration on astronomical sources and system gains
can be calibrated using the internal polarized noise generator on the S-Band
receiver. Since the \mmska\ has {linear polarizations}, calibrating
linearly polarized signal is expected to be challenging. However, the low
on-axis cross-coupling of $<-35$\,dB to circular polarization will reduce the
challenge of calibrating linear polarizations. The exact performance of \mmska\
will be {tested during its commissioning}.

\section{Advantages of a new S-Band survey} \label{sec:sband}

In this section, we discuss the main advantages of having broad bandwidth
spectro-polarimetric data in S-Band along with C-Band data from the C-BASS for
constraining the polarized synchrotron emission. We use additional lower and
higher frequency data, at 408~MHz and around 10~GHz, respectively, for
constraining the total synchrotron and free--free emissions.

\subsection{Recovering synchrotron and free--free emission foregrounds} \label{sec:model_total}

Following the motivation discussed in Section~\ref{sec:synchrotron}, we adopt
Eq.~\eqref{eq:model} to model the total intensity radio continuum emission
below 10~GHz in our Galaxy, i.e., as a sum of synchrotron emission ($T_{\rm
B,syn}$) and an optically thin free--free emission ($T_{\rm B, ff}$) --- so far
only used in the characterization of other galaxies' radio spectra \citep[see,
e.g.,][]{palad09, tabat17, klein18}. Since sharp spectral steepening is
somewhat unlikely in galactic ISM, we assume a fixed $\nu_{\rm c} \gg 10$ GHz,
which leads to setting the exponential-term in Eq.~\eqref{eq:model} to unity.
It is obvious that with either $\nu_{\rm br} \rightarrow \infty$ or $\gamma =
0$ we arrive back at an addition of two simple power-law spectra.

To assess the importance of having new broad-bandwidth data in the S-Band, here
we simulate a synthetic radio continuum spectrum and apply Eq.~\eqref{eq:model}
to constrain the contribution of synchrotron and free--free emission components
to the CMB. In the left-hand panel of Fig.~\ref{fig:spectra}, we generate a
model of the total radio continuum emission (shown as the solid black line) in
a $1\times1$ deg$^2$ pixel with a curved synchrotron emission component (shown
as the red dashed line) and an optically thin free--free emission component
(shown as the green dot-dashed line). The synchrotron emission has spectral
index $\bnt=-2.8$ and a break at $\nu_{\rm br} = 4\,\text{GHz}$, similar to
values found in \citet{klein18}. The spectral curvature $\gamma$ was fixed at
$\gamma=1.5$ --- a somewhat strong break, yet illustrative for our purpose. The
synchrotron brightness is normalized to 5 K at 1 GHz (a somewhat bright region
in the sky) for a $1\times 1$ deg$^2$ patch of sky, roughly corresponding to
the angular resolution of the \mmska.  The free--free emission component is
normalized to 10\% of the total emission at 1~GHz, a typical value observed in
spatially resolved nearby galaxies \citep{basu12a}. The fictive measurements
are performed at C-Band (4.5 to 5.5~GHz) divided into 128 channels using the
specifications of the Southern-sky survey of C-BASS listed in table~2 of
\citet{jones18}, and at S-Band (1.7 to 3.5~GHz), using the specifications of
\mmska\,listed in Table~\ref{tab:specs} with 200 channels. In addition, we
include a data point at Haslam's 408\,MHz with an uncertainty of 1~K and one
from a QUIJOTE- or GreenPol-like experiment at 10\,GHz with an uncertainty of
25~$\umu$K, consistent with their specifications (see also
Section~\ref{sec:quijote-south}).

We generated 500 data realizations with the input model, Eq.~\eqref{eq:model},
wherein the data points in each case were randomly drawn from a Gaussian noise
distribution. The noise for each data point includes: respective survey
sensitivity, confusion noise and a calibration error of 2\% was added to
highlight the robustness of our results in light of real data. The yellow and
grey lines in Fig.~\ref{fig:spectra} (left panel) show the result of fitting
\textit{with} and \textit{without} the S-Band data, respectively. The red and
the green lines show the recovered synchrotron and free--free emissions,
respectively, when S-Band data are included. This demonstrates the possibility
of separating the synchrotron and free--free emission components using
multi-frequency radio continuum data alone.

In the right-hand panel of Fig.~\ref{fig:spectra} we show the extrapolations of
the recovered spectra in the vicinity of 80\,GHz, relevant for CMB experiments.
For a comparison, the typical contribution of other foreground components, AME
and thermal dust, and the expected rms brightness temperature fluctuations of
the CMB are plotted as the different solid lines. The parameters describing the
AME and the thermal dust emission are taken from table 5 of \citet{jones18} and
are representative of a $1\times 1$ deg$^2$ patch of the sky where the
foregrounds are significant, i.e., for $|b| \lesssim 20^\circ$. It should be
noted that while the fits `with S-Band' follow the true signal closely, those
`without S-Band' are distributed bi-modally around the true values with biases
on both sides.\footnote{Given the steep spectrum of the synchrotron emission
above $\sim10\,\text{GHz}$, the total emission at 80\,GHz is essentially only
free--free. It is the bias in the free--free emission seen in the
right-hand-side panel.} Further note that, we have not fitted the other
foreground components and they are shown only for reference. In this example,
the dispersion of the brightness temperature ($\sigma_{80}$) of the recovered
synchrotron + free--free emission around the expected value at 80 GHz decreases
from $\sim25\,\umu$K without the S-Band data to $<10\,\umu$K by including the
S-Band data.

\begin{figure} 
\centering 
\begin{tabular}{c} 
{\mbox{\includegraphics[width=8cm,trim=0.5cm 2.8cm 0.5cm 3cm,clip]{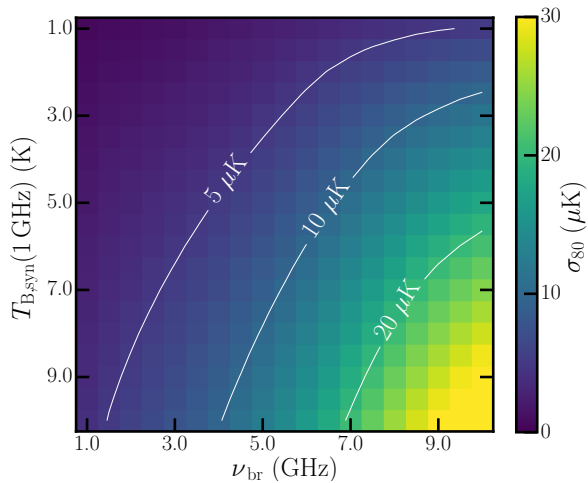}}}\\
\end{tabular}
\caption{Dispersion {$\sigma_{80}$} of model fits using
Eq.~\eqref{eq:model} around the expected synchrotron + free--free emission at
80 GHz for various choices of $\nu_{\rm br}$ and synchrotron normalizations
($T_{\rm B, syn}$) at 1 GHz for the case when S-Band data are included.}
\label{fig:spectra_disp} 
\end{figure}

In Fig.~\ref{fig:spectra_disp}, we show $\sigma_{80}$ as colour for various
values of break frequencies $\nu_{\rm br}$ and synchrotron normalizations at
1~GHz, $T_{\rm B, syn}({\rm 1\,GHz})$. For the typical range of values both
these parameters can have, synchrotron + free--free emission can be recovered
with rms better than $30\,\umu$K at 80 GHz. Low brightness temperature, $T_{\rm
B, syn}({\rm 1\,GHz}) < 5$~K, typically represent regions of high Galactic
latitude for which we find the residuals to be $\lesssim5\,\umu$K. It is in
these regions of the sky where $\nu_{\rm br}$ and $\nu_{\rm c}$ are expected to
be between 2 and 10 GHz. Therefore, new broadband measurements between 1.7 and
3.5~GHz with the \mmska\ will be important in constraining the synchrotron
spectrum also at high Galactic latitudes.

\begin{figure*} 
\begin{centering}
\begin{tabular}{cc}
\includegraphics[width=8.0cm]{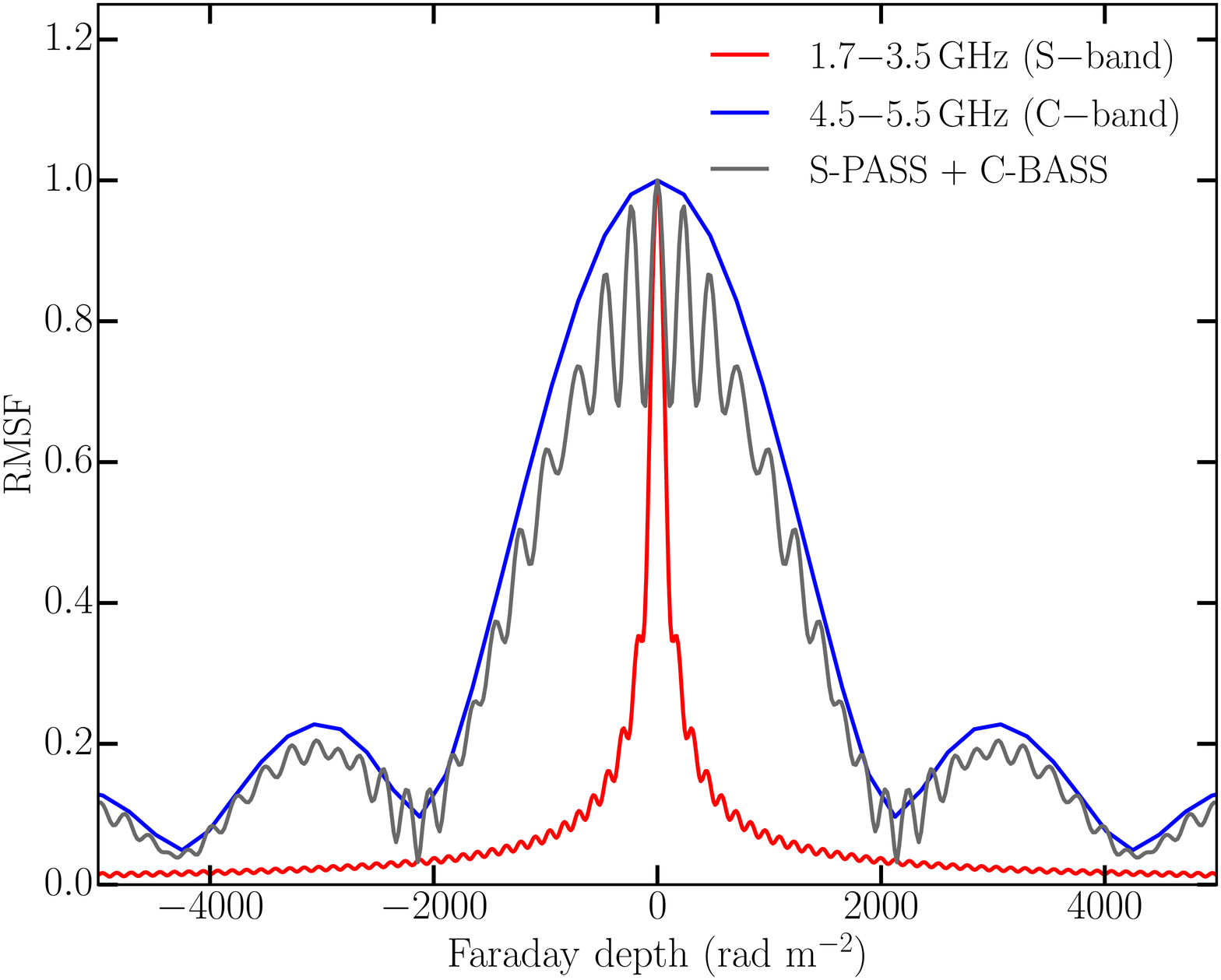} &
\includegraphics[width=8.0cm]{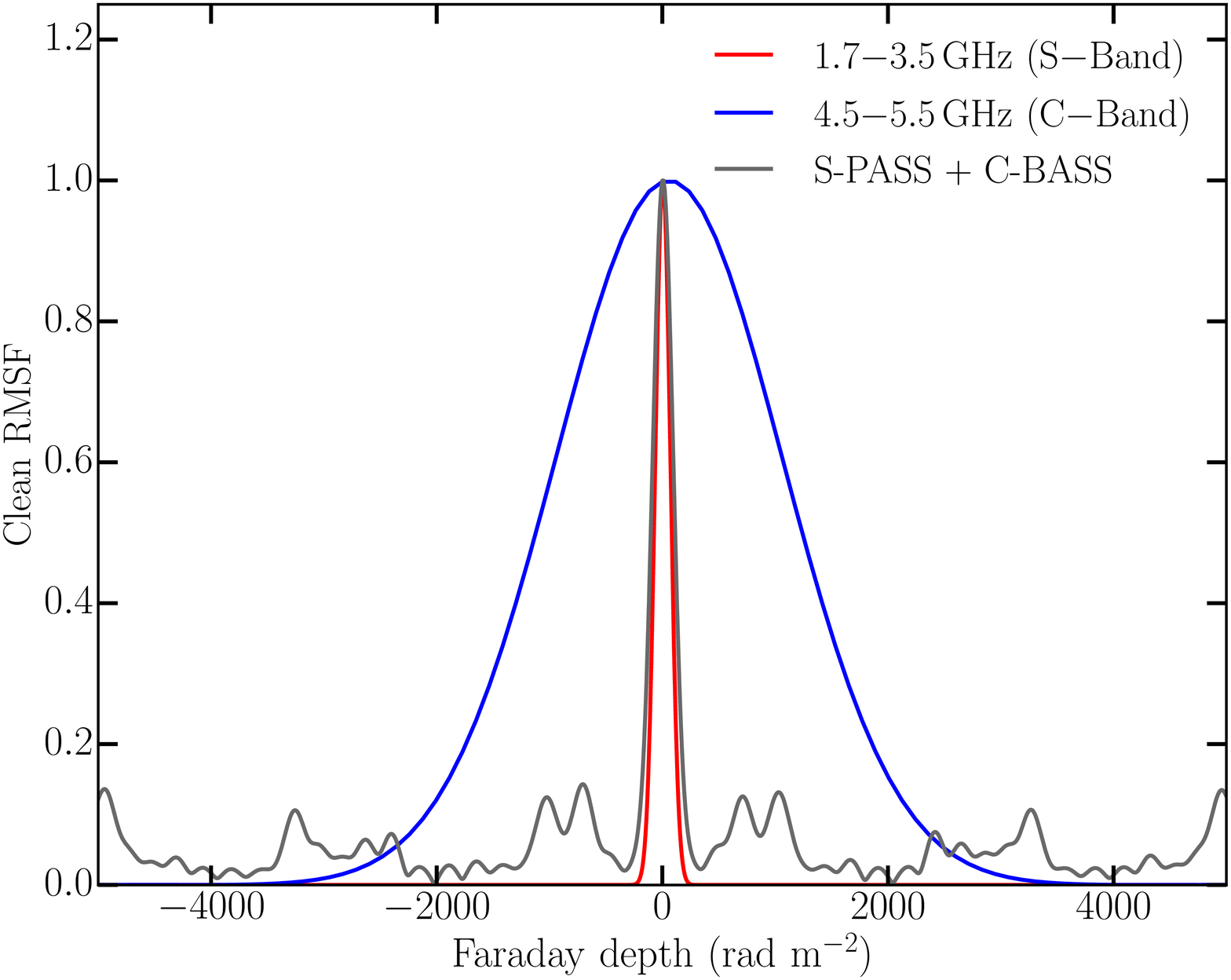}\\
\end{tabular}
\end{centering} 
\caption{{\it Left}: Rotation measure spread function (RMSF) for the S-Band
frequency coverage of \mmska\,is shown in red and has FWHM $\approx140\,\rm
rad\,m^{-2}$. RMSF with FWHM $\approx2400\,\rm rad\,m^{-2}$ for C-Band
frequency coverage of C-BASS is shown in blue. The grey RMSF is for a
combination of S-PASS and C-BASS frequencies with the same frequency weighting.
{\it Right}: Deconvolved (or cleaned) RMSFs for the corresponding frequency
coverages shown in the left-hand panel.} 
\label{fig:rmsf}
\end{figure*}

In bright regions with $\nu_{\rm br}$ lying above $\sim 7$~GHz, i.e., the
yellow region in the bottom right part of Fig.~\ref{fig:spectra_disp}, the
residuals increase to more than $20\,\umu$K. The CREs in the bright regions of
our Galaxy, i.e., close to the mid-plane region, are likely to be young and
$\nu_{\rm br}$ can lie above 10 GHz (see Section~\ref{sec:synchrotron}), even
up to 100~GHz for CREs accelerated in supernovae remnants which exploded $<
10^5$ years ago. Thus, in order to pin down the synchrotron spectra in regions
where $\nu_{\rm br}$ lies above 10~GHz, additional data in the Ku-Band and
higher frequencies, in the range 10 to 40~GHz, will be useful, provided the
increased AME contribution at those frequencies is also modeled with sufficient
accuracy. Therefore, with the parametrization of the radio continuum emission
below 10 GHz by Eq.~\eqref{eq:model}, we expect to recover the foreground
synchrotron + free--free emission around 80 GHz with rms $\lesssim 10\,\umu$K
per 1 deg$^2$ in most parts of the sky.

\subsection{Recovering the polarized synchrotron foreground}
\label{sec:model_pol}

Current, large sky-area surveys for measuring the polarized synchrotron
foreground using broad bandwidths, are either performed at too low frequencies
or at too high frequencies, severely limiting the scope of robustly determining
the effects of Faraday depolarization and applying Stokes $Q$, $U$ fitting
{(see Appendix~\ref{sec:survey})}. A spectro-polarimetric survey in
S-Band using the \mmska\ will provide the perfect opportunity to bridge the
large frequency gap between the C-BASS and the existing S-PASS. This opens up
the possibility to apply the technique of RM-synthesis \citep{brent05} and
directly produce a Galactic Faraday depth map without relying on Bayesian
techniques applied to Faraday depths measured for discrete extragalactic
sources \citep{opper12, opper15}.

The 1.7 to 3.5 GHz frequency coverage of \mmska\,will provide high Faraday
depth resolution with rotation measure spread function (RMSF) FWHM of
$\approx140\,\rm rad\,m^{-2}$ in contrast to $\approx2400\,\rm rad\,m^{-2}$ for
C-BASS. The RMSFs for the two frequency coverages, and for comparison, a
combination of the C-BASS and the S-PASS frequencies are shown in the left-hand
panel of Fig.~\ref{fig:rmsf}. In the right-hand panel of Fig.~\ref{fig:rmsf},
the corresponding deconvolved (or cleaned) RMSFs are shown. Note that, for the
combined S-PASS+C-BASS data, residual sidelobes at 15 per cent level are seen.
The accuracy ($\delta\textrm{FD}$) to which FD can be estimated is related to
the FWHM of the RMSF as, $\delta\textrm{FD} = \textrm{FWHM} / (2
\,\textrm{SNR})$, where SNR is the signal-to-noise ratio of the linearly
polarized intensity. This relation is applicable to the simple scenario when
there is a single FD component within the RMSF peak, or for a synchrotron
emitting and Faraday rotating medium which has $\rm FD \ll FWHM$ and/or $\rm
\sigma_{FD} \ll FWHM$. 

Since, the resolution of FD in C-Band is poor, C-BASS data alone would be
inadequate to distinguish complicated and/or wide Faraday depth structures
{in the Faraday depth spectrum originating due to Faraday
depolarization} (see Section~\ref{sec:depol}). Moreover, it also limits the
accuracy to which the polarization angle measured using the C-BASS data can be
corrected for Faraday rotation. Combining C-BASS and S-PASS data would improve
FD measurements for sources with simple Faraday depth spectra, e.g., at high
Galactic latitudes ($|b| \gtrsim 30^\circ$). But, sources extended in Faraday
depth would suffer from inaccurate flux recovery due to the 15 per cent
sidelobes of the deconvolved RMSF (see Fig.~\ref{fig:rmsf}). This will be
severe at $|b| \lesssim3 0^\circ$. With the availability of broadband data at
S-Band using the \mmska, we will be able to measure Faraday depths at better
than $\sim15\, \rm rad\,m^{-2}$ accuracy up to Faraday depths of $\pm 4\times
10^4\,\rm rad\,m^{-2}$ and be sensitive to extended Faraday depth structures up
to $\sim450\,\rm rad\,m^{-2}$. This would allow us to measure complicated
Faraday depth structures \citep{frick11} from the turbulent interstellar medium
at low Galactic latitudes and also produce a robust Southern-sky Faraday depth
map.

\begin{figure*} 
\begin{centering} 
\begin{tabular}{cc}
{\mbox{\includegraphics[width=8cm]{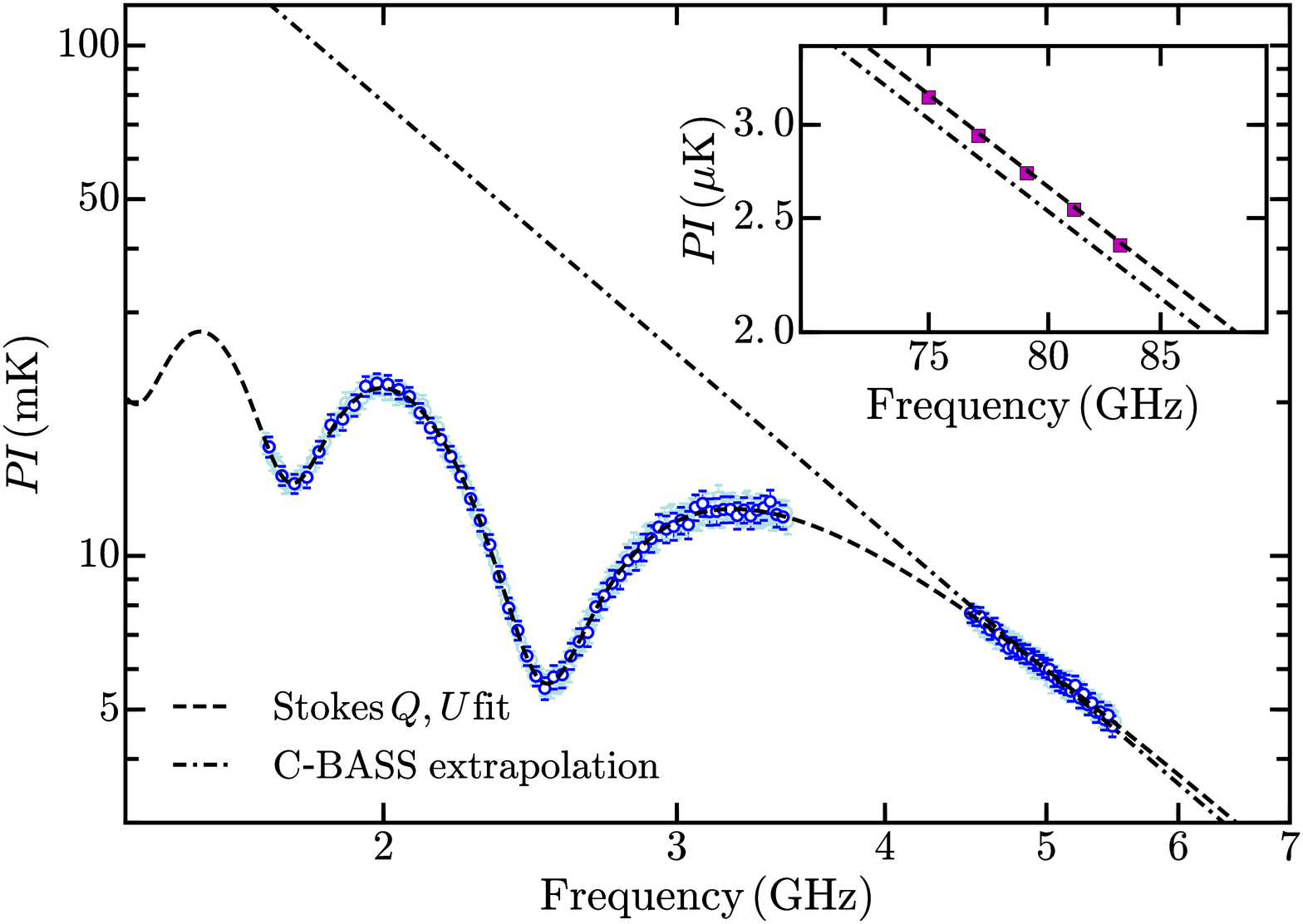}}} &
{\mbox{\includegraphics[width=8cm]{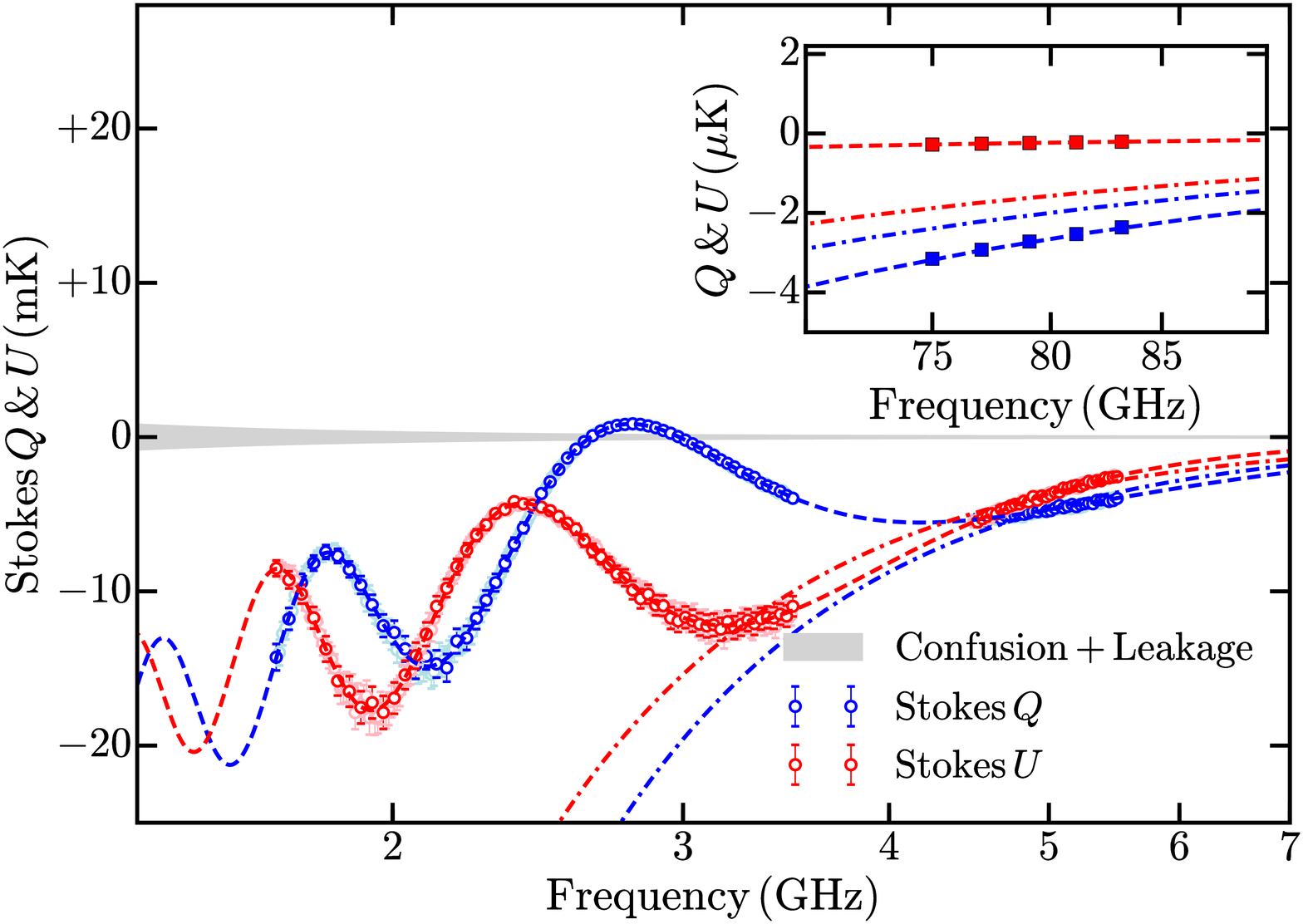}}} \\
\end{tabular} 
\end{centering} 
\caption{Left- and right-hand panels show the frequency spectrum of $PI$, and
Stokes $Q$ and $U$, respectively. The data points are generated from synthetic
observations of a MHD simulation of isothermal compressible turbulence in the
ISM. For clearer visual appearance, we have highlighted every fourth data point
and the errorbars show $3\sigma$. The dashed lines show the best fit model
using Stokes $Q$, $U$ fitting (listed in Table~\ref{tab:qufit}). The dot-dashed
line show a simple power-law extrapolation of the polarized intensities
measured with the C-Band data only. The shaded area represents confusion noise
scaled to 1 degree angular resolution from polarized sources and an
instrumental polarization leakage of 0.2\%, added in quadrature. The insets
show zoomed-in {spectra} around 80~GHz. These data were not used
while performing the fit and should only serve as guides to evaluate the
accuracy of the Stokes $Q,U$ fitting technique.} 
\label{fig:qufit} 
\end{figure*}

\subsubsection{Regions of high Faraday depolarization} \label{sec:qufit}

Note that, RM-synthesis is an important tool to recover polarized signals that
are otherwise too faint to be detected at the sensitivity level of individual
narrow channels and/or could suffer from significant depolarization when
averaged over large bandwidths. However, interpreting complicated Faraday depth
spectra to recover the intrinsic polarization angle and fractional polarization
is difficult \citep[see e.g.,][]{ander16}. The technique of direct fitting of
the observed Stokes $Q$ and $U$ spectra with analytic models of turbulent
magneto-ionic medium offers a better solution to study a Faraday complex
medium. To demonstrate the power of Stokes $Q$, $U$ fitting, here we show an
example using a synthetic polarization spectrum constructed from a
magnetohydrodynamic (MHD) simulation of turbulent ISM. 

Polarization observations of the Galactic plane have revealed the turbulence in
the warm diffuse phase of interstellar gas to be low sonic Mach number through
comparisons with MHD simulations of isothermal compressible turbulence
\citep{burkh09, gaens11, burkh12, herron2016}. The sub-sonic nature of ISM
turbulence in the warm/hot phases has also been confirmed through Galactic
H{\sc i} 21-cm observations \citep{burkh2010, koley19}. Here, we use the same
MHD simulation of a $512 \times 512 \times 512\,\,\rm pc^{3}$ volume having
mesh resolution of $1 \times 1 \times 1\,\,\rm pc^{3}$ from \citet{burkh09,
burkh13}.\footnote{For details on the numerical setup, see \cite{kowal07},
\cite{burkh09} and \cite{bialy17}.} The simulation consists of a background
magnetic field of strength $10\,\umu$G in the plane of the sky, and the ratio
of ordered to turbulent magnetic field strengths is $\sim2$. The median
electron density ($n_{\rm e}$) in the simulation volume is $0.1\,\rm cm^{-3}$
and has a maximum value of $\sim3.5\,\rm cm^{-3}$. Faraday depth integrated
along the line of sight varies between $-350$ and $+350$~rad\,m$^{-2}$ per $1
\times 1$~pc$^2$ pixel. Such values of $n_{\rm e}$ and FD are typically
encountered in the thin disc of the Milky Way, roughly within $|b| < 20^\circ$
\citep[see][respectively]{ne2001II, opper12}.  

Synthetic broadband observations of the linearly polarized synchrotron emission
including the effects of Faraday rotation were generated for the above
mentioned simulation (A. Basu et al. in preparation). The Stokes $Q$ and $U$
parameters were calculated for a 512 pc long line of sight and averaged over
$30 \times 30\,\,\rm pc^2$. This set-up corresponds roughly to a $1\times
1\rm\,\,deg^2$ patch on the sky at a distance of 10 kpc, the typical distance
in the Galaxy and roughly matched to the angular resolution of a survey with
the \mmska.

We generated values for Stokes $Q$ and $U$ in the frequency ranges, 1.7 to 3.5
GHz (S-Band covered by \mmska) divided into 200 frequency channels, 4.5 to 5.5
GHz (C-Band covered by C-BASS in the South) divided into 128 frequency channels
and some around 80 GHz, i.e., a representative high frequency near which the
CMB polarized foreground is to be estimated. Stokes $Q,U$ values near 80~GHz
were computed to compare results of extrapolation and were not used for
fitting. {Sensitivity of the respective surveys,} a calibration error
of 2\% and confusion noise from polarized sources were added {as
Gaussian noise} to the Stokes $Q$ and $U$ data in the S- and C-Bands. The total
synchrotron emission was normalized to 1 K at 1 GHz for the 1 deg$^2$ patch of
sky, a somewhat faint region to assess sensitivity to Faraday depolarization
features in S-Band.

\begin{table} 
\centering
\caption{Best fit parameters for the three internal Faraday dispersion (IFD) 
components used as the model for Stokes $Q,U$ fitting in Figs.~\ref{fig:qufit}
and \ref{fig:polangle}.}
\centering
  \begin{tabular}{@{}lccc@{}}
 \hline
Component  &  1 & 2 &  3 \\
Model	  & IFD & IFD & IFD \\
 \hline
	  $p_0$ & $0.45\pm0.01$ & $0.11\pm0.04$ & $0.11\pm0.04$ \\
	  {$\rm FD_{em}$} ($\rm rad\,m^{-2}$) & $203.1\pm0.7$ & $71.7\pm42.1$ &$9.7\pm14.9$ \\
	  $\sigma_{\rm FD}$ ($\rm rad\,m^{-2}$)  &$14.2\pm0.4$  & $55.9\pm18.2$ & $21.5\pm5.5$ \\
	  $\theta_0$ (degree) & $86.6\pm1.0$&  $-48.1\pm14.3$ & $88.6\pm15.2$ \\
\hline
\end{tabular}
\label{tab:qufit}
\end{table}

{The data points in} Fig.~\ref{fig:qufit} shows the synthetic
frequency spectrum of the linearly polarized intensity ($PI$) in the left-hand
panel and, Stokes $Q$ and $U$ parameters in the right-hand panel. Note that
large $\lambda^2$ coverage is essential for robust Stokes $Q, U$ fitting,
otherwise the fitted models and their parameters could be degenerate giving
rise to systematic uncertainties.  By combining S-Band data to C-Band data, the
$\lambda^2$ coverage increases by a factor of about 20, i.e., from $\Delta
\lambda^2 \approx 14.5\,{\rm cm}^2$ in C-Band for the C-BASS to $\Delta
\lambda^2 \approx 280\,{\rm cm}^2$. We therefore combined S- and C-Band to
perform Stokes $Q, U$ fitting. 

{For this simulation, we are always looking at a Faraday rotating
medium which is simultaneously synchrotron emitting. Also, there is no purely
Faraday rotating medium in the foreground of the simulation volume. Therefore,
in our case FD and $\rm FD_{em}$ are equivalent (see Section~\ref{sec:depol}).
For performing Stokes $Q,U$ fitting, we have included various linear
combinations of models of magneto-ionic media discussed in
Section~\ref{sec:depol}, including purely Faraday rotating media, and used the
corrected Akaike information criteria \citep{hurvi89, cavan97} to choose the
preferred model.}

{For the example shown in Fig.~\ref{fig:qufit}}, the best fit model
was obtained for a linear combination of three internal Faraday dispersion
components (Eq.~\ref{eq:int_disp}) shown as the dashed lines in
Fig.~\ref{fig:qufit}. Best fit parameters are presented in
Table~\ref{tab:qufit}. The dot-dashed line in the Fig.~\ref{fig:qufit} shows a
simple power-law extrapolation of the linearly polarized intensity measured at
C-Band only. This is equivalent to assuming that the frequency-dependent
Faraday depolarization at C-Band is negligible and the polarized intensity has
the same spectral shape as that of the total synchrotron intensity \citep[see
e.g.,][]{planck2015X}, in this example a power-law with $\beta_{\rm nt} =
-2.8$. A cursory look at the recovered Stokes $Q$ and $U$ parameters at C-Band
using both the approaches appears to be in agreement within the errors.
However, the technique of Stokes $Q,U$ fitting captures the depolarization
features imprinted on the polarized intensity spectrum at lower frequencies
much better than the power-law approach. This information is crucial to
estimate the contribution of the polarized synchrotron emission to the CMB
foreground with cosmological precision.

The insets in Fig.~\ref{fig:qufit} show the extrapolations around 80 GHz. The
square data points show the expected polarized signal and were not included
while fitting. The dashed and dot-dashed lines are the respective predictions
of the linearly polarized synchrotron intensity around 80 GHz using Stokes
$Q,U$ fitting and assuming power-law spectrum from C-Band only measurements.
Although the power-law extrapolation from the C-Band measurements well
represents the polarized quantities around 80 GHz within sub-$\umu$K accuracy,
the recovered Stokes $Q$ and $U$ parameters are significantly off. This is
critical for measuring the polarization angle of the synchrotron emission. In
order to accurately decompose the polarized CMB into \textit{E}- and
\textit{B}-modes, errors in the polarization angle measurement of the
foreground synchrotron emission could result in spurious mixing of the modes.

\begin{figure} 
\begin{centering}
\begin{tabular}{c}
{\mbox{\includegraphics[width=8cm]{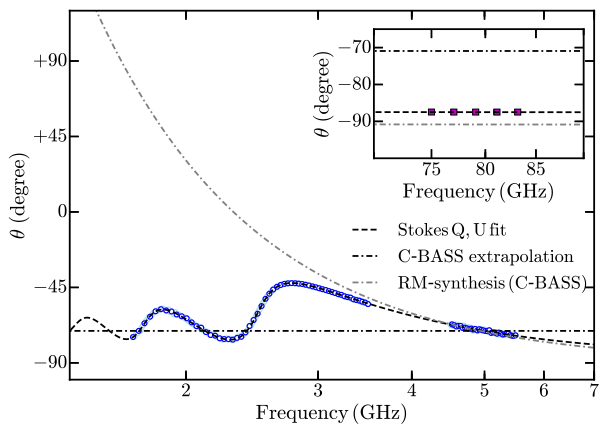}}} \\ 
\end{tabular}
\end{centering} 
\caption{Variation of the polarization angle with frequency for the polarized
emission shown in Fig.~\ref{fig:qufit}. Every fourth data point is highlighted
for visual clarity. The dashed line shows the result of Stokes $Q$, $U$
fitting, the black dot-dashed line shows the simple extrapolation of the
polarization angle using the measurement at C-Band and the grey dot-dashed line
shows the result of RM-synthesis applied to the C-Band only measurements.}
\label{fig:polangle} 
\end{figure}

In Fig.~\ref{fig:polangle}, we show the polarization angle ($\theta$) computed
from the Stokes $Q$ and $U$ parameters in Fig.~\ref{fig:qufit} (right-hand
panel) as a function of frequency. The result of Stokes $Q,U$ fitting is shown
as the dashed line, simple extrapolation of the polarization angle measured by
averaging over C-Band\footnote{To measure frequency averaged polarization angle,
we first averaged the Stokes $Q$ and $U$ intensities over the C-Band and then
computed the polarization angle.} is shown as the black dot-dashed line and
from RM-synthesis applied to the C-Band measurements is shown as the grey
dot-dashed line. The inset shows the results of extrapolation in the vicinity
of 80 GHz. It is obvious that a direct extrapolation from the C-Band averaged
polarization angle does not recover the polarization angle around 80 GHz, and
the estimated angles are off by more than $15^\circ$. Even by applying the
technique of RM-synthesis to the C-Band only data, the recovered polarization
angle is a couple of degrees off from the expected angle. This is mainly due
the fact that, low Faraday depth resolution of $\sim$2500 $\rm rad\,m^{-2}$ at
C-Band is insensitive to the complicated Faraday depth structures in the FD
spectrum. In contrast, with the availability of S-Band data, Stokes $Q,U$
fitting recovers the polarization angle at a fraction of degree accuracy. We
however note that the accuracy of polarization angle measurement is limited by
the accuracy to which the polarization angle of calibrators are typically
measured and is about 1--2 degrees. Although Stokes $Q, U$ fitting can recover
the angle within fraction of a degree, systematic angle offsets due to
uncertainty in the polarization angle of the calibrators cannot be excluded.

\begin{figure} 
\begin{centering} 
\begin{tabular}{c}
{\mbox{~~~\includegraphics[width=7.5cm]{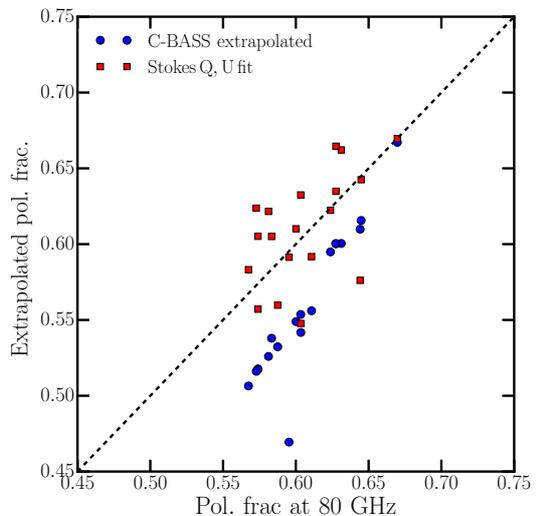}}} \\
\end{tabular} 
\end{centering} 
\caption{Extrapolated polarization fraction at 80 GHz as function of expected
polarization fraction at 80 GHz determined from synthetic observations of MHD
simulations. The red points are for the polarization fraction determined using
the technique of Stokes $Q, U$ fitting applied to combined S-Band and C-Band
data. The blue points are for simple power-law extrapolation from observations
at C-Band only. The errors on the data are smaller than the scatter.} 
\label{fig:forecast} 
\end{figure}

In Fig.~\ref{fig:forecast} we show the extrapolated fractional polarization at
80 GHz using Stokes $Q, U$ fitting to the combined S- and C-Band measurements
as red points, and simple power-law extrapolation of the C-Band only
measurements as the blue points, as a function of the expected fractional
polarization at 80 GHz for a random set of sightlines. In this example, the
C-Band only extrapolation suffers from systematic uncertainties up to $\sim
-20\%$ with a mean of $-10\%$, whereas the uncertainty using Stokes $Q, U$
fitting is mostly statistical with up to $\sim \pm 10\%$ and has a dispersion
of 5\% around the expected fractional polarization at 80~GHz. In this example,
the intrinsic fractional polarization ($p_0$) and thereby the fractional
polarization near 80~GHz are high (see Fig.~\ref{fig:forecast}) due to the
relatively strong background {magnetic} field strength used in the
MHD simulation. A lower ratio of the strengths of ordered to turbulent
{magnetic} fields will lead to a lower $p_0$ for similar FD values
used here. This will not change the overall results presented in this section.

We would like to point out that, the nature of systematic deviation of the
C-Band extrapolated polarized emission discussed above might be a feature of
the MHD simulation used by us as an example. A different set of simulation
could lead to different systematics. This is being investigated with various
type of MHD simulations of the turbulent ISM (A. Basu et al. in preparation).
However, we do believe that using extrapolated C-Band only observations to
model the foreground polarized synchrotron emission, can introduce undesirable
systematics in the estimated polarized CMB map and therewith the polarized CMB
angular power spectrum.

\begin{figure} 
\begin{centering} 
\begin{tabular}{c}
{\mbox{\includegraphics[width=7.6cm]{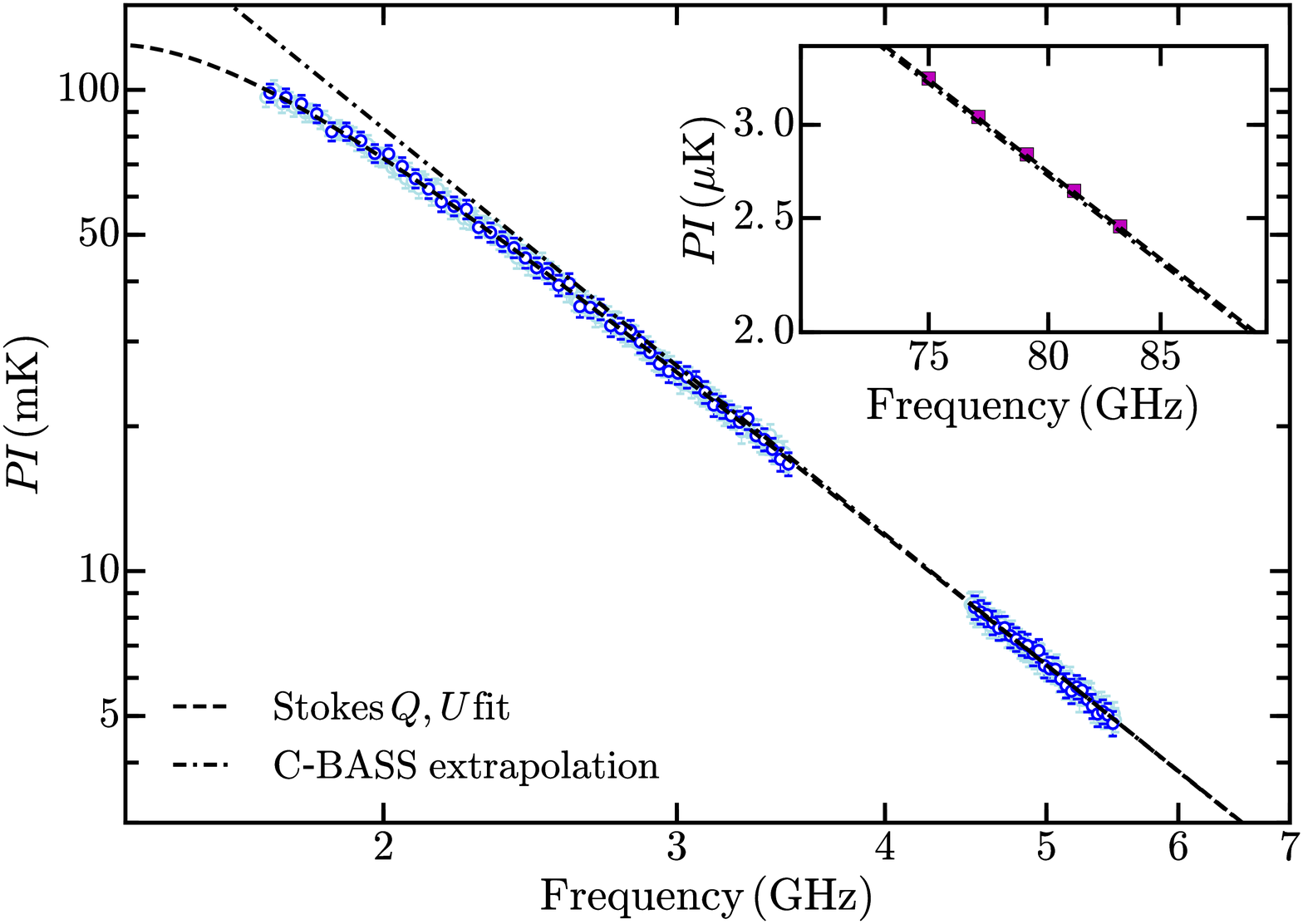}}} \\
{\mbox{\includegraphics[width=8cm]{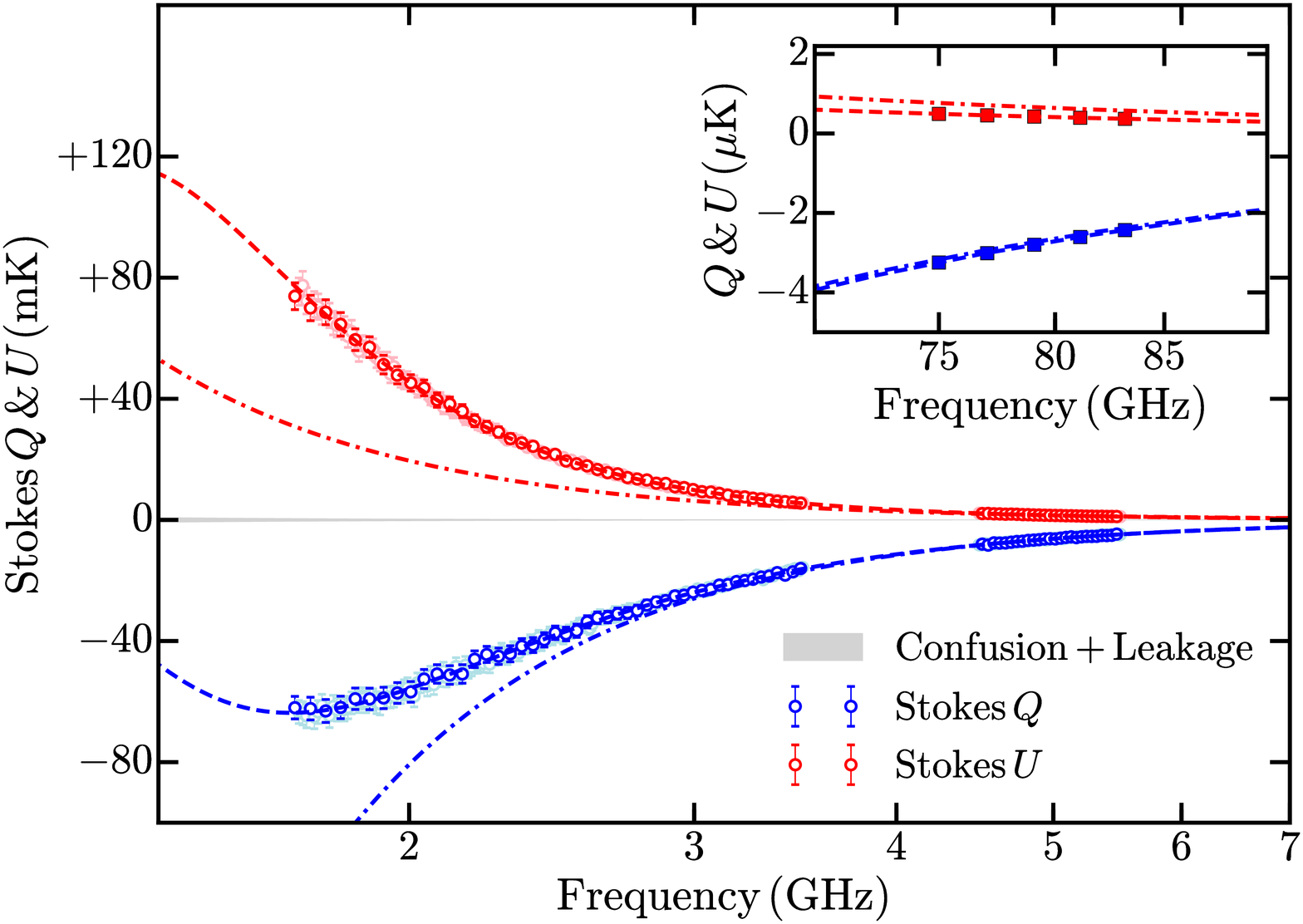}}} \\
{\mbox{\includegraphics[width=8cm]{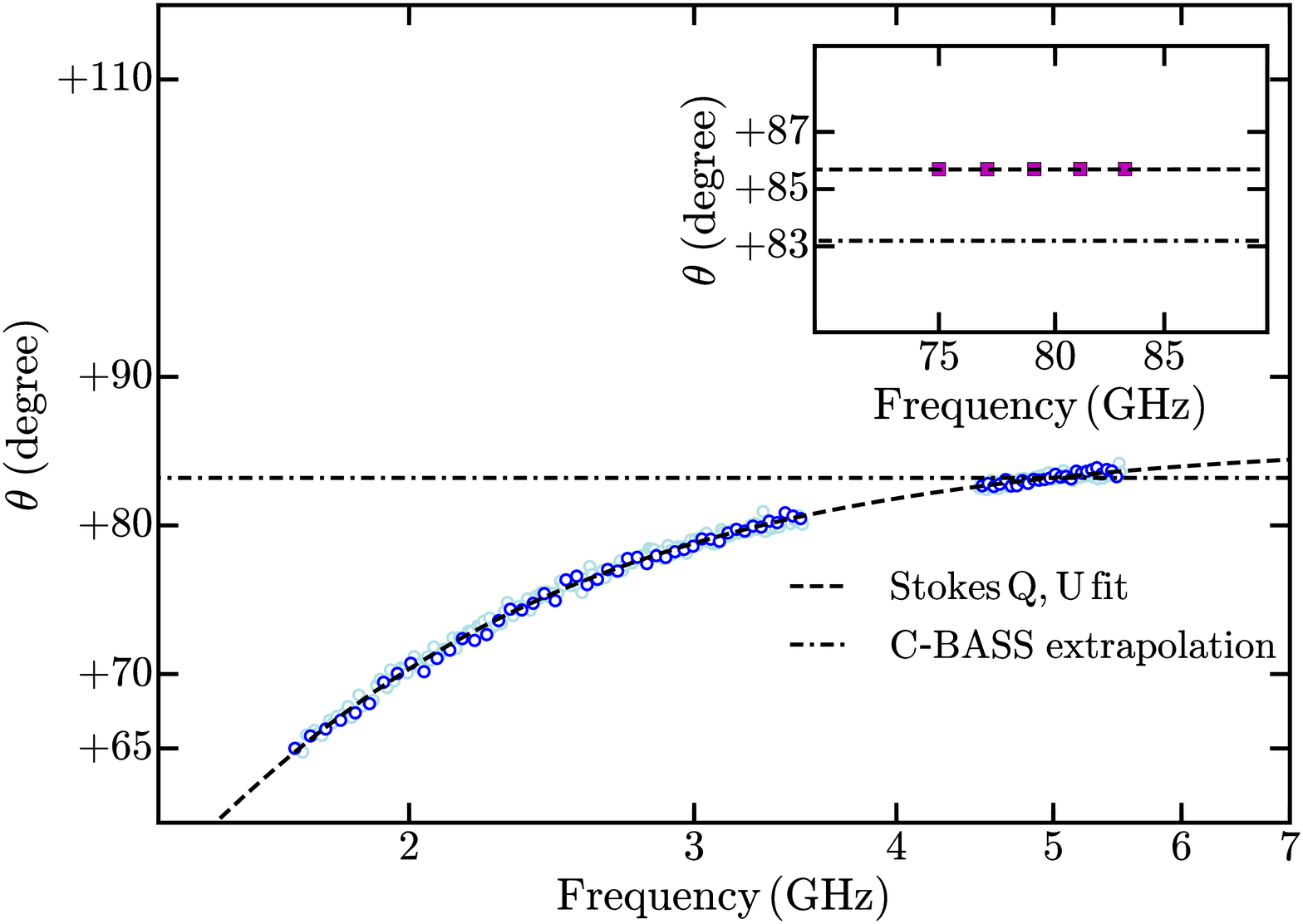}}} \\
\end{tabular} 
\end{centering} 
\caption{Example spectra of linear polarization parameters in a region where
effects of Faraday depolarization is low. Top-, middle- and bottom-panels show
the spectrum of $PI$, Stokes $Q$ and $U$, and polarization angle, respectively.
The points, lines and insets have the same meaning as the corresponding plots
in Figs.~\ref{fig:qufit} and \ref{fig:polangle}.} 
\label{fig:qufit_freeze} 
\end{figure}

\subsubsection{Regions of low Faraday depolarization}

The example polarization spectra presented in the previous section
show a region where frequency dependent Faraday depolarization is strong. In
this section we discuss the properties of polarization spectra for a region
where Faraday depolarization is less severe and representative of polarized
emission from high Galactic latitudes. Unfortunately, we do not have access to
MHD simulations for high galactic latitudes. Therefore, in order to mimic
polarized emission from such regions, we simply scaled up the simulation volume
used in the previous section. Magnetic field strengths were scaled using the
assumption of magnetic flux freezing and keeping the gas mass conserved. Thus,
for our scaling, $B/\rho^{2/3}$ is conserved, where $B$ is the magnetic field
strength and $\rho$ is the gas density.\footnote{Strictly speaking, this
approach does not represent the energetics and kinematics of actual physical
conditions at high galactic latitudes, but is sufficient to extract synthetic
polarization spectra with low Faraday depolarization.} Here, we scaled the
simulation volume to $1024\times1024\times1024$ pc$^3$, i.e., a factor of 8
increase in volume of the original simulation. The median $n_{\rm e}$ in this
case is 0.03~cm$^{-3}$ and FD varies between $-40$ and $+40$~rad~m$^{-2}$ per
$2 \times 2$~pc$^2$ pixel. This represents the mid-Galactic latitude region,
$30^\circ \lesssim |b| \lesssim 60^\circ$.

\begin{table} 
\centering
\caption{Best fit parameters for the two Burn slab components used as the model
for Stokes $Q,U$ fitting in Fig.~\ref{fig:qufit_freeze}}
\centering
  \begin{tabular}{@{}lcc@{}}
 \hline
Component  & 1 &  2  \\
Model	  & Burn & Burn \\
 \hline
	  $p_0$ & $0.43\pm0.02$ & $0.16\pm0.02$  \\
	  {$\rm FD_{em}$} ($\rm rad\,m^{-2}$) & $-32.3\pm0.7$ & $-2.2\pm2.7$  \\
	  $\theta_0$ (degree) & $83.6\pm0.8$&  $-88.7\pm1.8$  \\
\hline
\end{tabular}
\label{tab:qufit_freeze}
\end{table}

In Fig.~\ref{fig:qufit_freeze} we show the frequency spectrum of quantities
describing the linearly polarized emission averaged over a $30\times30$ pc$^2$
region. The total synchrotron brightness is normalized to 1~K at 1~GHz. In
this case, the best fit was obtained with {a linear combination of}
two Burn slab components (Eq.~\ref{eq:burn}) and the best fit parameters are
listed in Table~\ref{tab:qufit_freeze}. From the top-panel of the figure, it is
clear that both Stokes $Q,U$ fitting and simple power-law extrapolation from
C-Band measurements, recovers the polarized intensity near 80~GHz accurately.
However, the approach of simple power-law extrapolation gives rise to slight
systematic deviation in the extrapolated Stokes $Q$ and $U$ parameters as seen
in the inset of the middle-panel in Fig.~\ref{fig:qufit_freeze}. This results
in the polarization angle estimation near 80~GHz to be off by a few degrees
(see inset in the bottom-panel of Fig.~\ref{fig:qufit_freeze}). In fact, in
regions of low Faraday depolarization, performing Stokes $Q, U$ fitting by
combining C-BASS data with S-PASS will improve estimation of both polarized
intensity and polarization angle with accuracy similar to what we find by
combining broad-band S-Band data with C-BASS.

At high Galactic latitudes $|b| \gtrsim 60^\circ$, Faraday depolarization is
expected to be even lower, and therefore extrapolation of polarization
measurements in C-Band, e.g., from the C-BASS, by means of simple power-law,
could recover the foreground polarized synchrotron emission at CMB frequencies
quite well. Regions in the sky where such an extrapolation is applicable covers
limited area. This thereby limits the measurement of angular power spectra of
the polarized CMB at the largest angular scales. However, the method of Stokes
$Q,U$ fitting, by including S-Band measurements from the \mmska\ will give
better accuracy, almost over the entire Southern sky.

\section{Possibility of QUIJOTE-like counterpart for the Southern sky}
\label{sec:quijote-south}

Currently the QUIJOTE experiment is being performed at the Teide observatory in
Tenerife, Canary Islands to cover the Northern sky in the frequency
range 11 to 40 GHz which aims to measure the Galactic AME and polarized dust
emission (see Appendix~\ref{sec:quijote}). As mentioned in \citet{jones18}, 
plans to extend QUIJOTE to the Southern hemisphere is not yet funded. The
\mmska\ can also be equipped with Ku-band receivers operating in the frequency
range 12 to 18 GHz equipped with a GPU based backend. This frequency coverage
is similar to that of QUIJOTE's MFI and would be ideal for conducting a
QUIJOTE-like survey in the Southern hemisphere.

We however note that, since the \mmska\ was not originally designed for such an
experiment, it has a small field of view (7 to 4.6~arcmin) because of a
relatively large dish-size for a survey instrument at these frequencies (see
Table~\ref{tab:specs}). In order to achieve a sensitivity of $\sim25~\umu$K, as
targeted by the QUIJOTE experiment, $\sim$\,10 min per resolution element will
be required. This will lead to a total survey time of $\gtrsim\,2 \times
60\,000$ hours to cover the Southern sky with the full 6~GHz ($2\times3$~GHz)
bandwidth of the \mmska\ in the Ku-Band. However, a smaller area on the sky,
for example, the area of 2500 deg$^2$ covered by the South Pole Telescope
\citep{georg15} can be surveyed within a feasible time-scale of $\sim$10\,000
hours using the \mmska. A study of the feasibility of such a survey by means of
the \mmska\ is in progress and its results will be reported elsewhere.

\section{Discussion, outlook and summary} \label{sec:summary}

The Galactic synchrotron and free--free emissions contribute significantly to
the low-frequency CMB intensity foreground. Their separation is usually
performed either via parametric foreground fits in each pixel
\citep{planck2015X}, or via non-parametric template fits \citep{benne13}. The
success of the latter, among others, depends on the accuracy with which
foreground templates can be obtained. For instance, H$\alpha$ maps, which serve
as template for free--free emission, suffer from significant absorption by
dust, especially around the Galactic plane up to $\pm15^\circ$ in latitude, for
which corrections have to be performed \citep[see, e.g.,][]{dicki03}. The
\citet{hasla82} all-sky survey at 408 MHz is considered as a template for the
synchrotron emission \citep{planck2015XXV}. Alternatively, a combined
parametric fit for all components is common practice (see, e.g., the
\texttt{Commander} method in \citealt{planck2018IV}.). Here, however,
degeneracies with other low-frequency foregrounds such as AME are inevitable,
and due to their model uncertainties misidentification are likely. 

Parametric pixel fits require assumptions on the spectral shape of the
foregrounds considered, and possible degeneracies among different foregrounds,
especially \textit{above} $\sim$10~GHz, must be controlled. Currently, due to
the lack of suitable large sky-area surveys covering multiple frequencies below
10~GHz, modelling of the synchrotron spectrum is limited to minimal
parametrization. For example, to estimate the contribution of the Galactic
synchrotron emission at CMB frequencies, it is modeled as a perfect power-law
with spectral index $\beta_{\rm nt}$ or with a constant spectral curvature over
the entire sky by modifying $\bnt$ as $\beta_{\rm nt}+C\log(\nu/\nu_{\rm br})$
\citep[see e.g.,][]{kogut12}. Here, $C$ is a curvature parameter. Such
parametrizations do not adequately describe the physical nature of the
synchrotron spectrum due to energy losses and propagation of the synchrotron
emitting CREs discussed in Section~\ref{sec:synchrotron}.

Broad-bandwidth data at S-Band, along with other on-going surveys at slightly
higher frequencies, like the C-BASS and QUIJOTE, provides the opportunity to
describe the radio continuum spectrum with physically motivated parametrization
of the total synchrotron emission. We hereto suggest the parametrization
presented in Section~\ref{sec:synchrotron} and test its feasibility using
simulated data in Section~\ref{sec:model_total}.  More generally, we propose
that large frequency coverage at frequencies \textit{below} $\sim$10~GHz has
the advantage of constraining both synchrotron and free-free emission directly
from the Galactic radio continuum spectrum alone. This thereby avoids
degeneracies with the CMB itself, AME, CO emission and thermal dust, and
circumvents uncertainties arising from templates. 

In Section~\ref{sec:model_total} we demonstrated the ability to constrain
physical models of radio emission with the help of \mmska. We expect that a
single, 300-hour S-Band survey of the Southern sky will provide a confusion
limited intensity measurement resolved into 21\,000 independent pixels with 200
frequency channels in each. Each single 300-hour survey when combined with
surveys at higher frequencies will allow extrapolation of synchrotron and
free--free emission to CMB frequencies with a rms brightness temperature
uncertainty of less than $10\,\umu$K in most parts of the Galaxy.  Survey at
S-Band with the \mmska\ will be specially helpful to model curvature in
synchrotron emission spectrum at high Galactic latitudes, $|b|\gtrsim
30^\circ$. For $|b| \lesssim 30^\circ$, our proposed method of simultaneous
modeling of the synchrotron and free--free emissions will benefit from
additional data in the frequency range 10 to 40~GHz. This in turn will lead to
an increase of the fraction of sky that can be used for CMB analysis and thus
help in reducing the limitations due to cosmic variance.

The full power of the planned S-Band survey however is revealed when we turn to
the study of the polarized synchrotron emission with the \mmska, as discussed
in detail in Section~\ref{sec:model_pol}. Repeated 300 hours S-Band surveys are
planned to eventually build up polarized maps with 0.3 mK sensitivity across
200 frequency channels spread over the S-Band.

From Figs.~\ref{fig:qufit} and \ref{fig:polangle}, it is clear that the
technique of Stokes $Q, U$ fitting is a powerful tool to estimate the linearly
polarized emission, including the intrinsic polarization angle, of the
foreground synchrotron emission which is often affected by Faraday
depolarization. S-Band coverage using the \mmska\ is a valuable addition to
applying this technique and does not suffer from model degeneracies as compared
to higher frequency observations around C-Band. In fact, a deeper understanding
of the foreground polarized emission through broad-band spectro-polarimetry is
imperative to disentangle between local and cosmological structures. For
example, residual signal of Galactic Loop I in the CMB's \textit{E}-mode map
has been hinted at in \citet{liu18}. The example MHD simulation we have used in
Section~\ref{sec:qufit} is representative of the ISM condition around the
Galactic plane region extending to $\sim\pm20^\circ$ in latitude and perhaps
for the large-scale polarized loops and spurs extending to even higher/lower
latitudes.  As these regions are affected by severe Faraday depolarizations,
they are usually masked from the polarized CMB angular power spectrum analysis.
Stokes $Q,U$ fitting is a powerful technique to model Faraday depolarization
and thereby will increase the usable sky-area drastically. 

Applying Stokes $Q,U$ fitting to the combined S-Band data from the \mmska\ and
C-Band data from the C-BASS, we expect to recover the polarization fraction of
the synchrotron emission near 80~GHz with a dispersion of 5\% and achieve an
accuracy better than a degree in estimating the angle of the linearly polarized
synchrotron emission, even at $|b| \lesssim 30^\circ$. This is essential for
reducing cosmic variance on the largest angular scales where the
next-generation of CMB experiments are targeting to detect the imprints of
primordial gravitational waves in the \textit{B}-mode and effects of
reionization history both in the \textit{E}- and \textit{B}-mode. 

Future CMB space missions are aiming to constrain the $B$-mode peak for
tensor-to-scalar perturbation power ratio $r$ with sensitivity $\delta r <
0.001$ \citep{suzuki18}. This corresponds to rms of roughly 10~nK for the
$B$-mode signal on 1 degree angular scales. In order to achieve these
sensitivities, the polarized synchrotron emission in the foreground should be
constrained with rms better than 5~nK at 100~GHz, i.e., better than $\sim10$~nK
at 80~GHz. Now, with 5\% dispersion on polarization fraction estimation using
Stokes $Q, U$ fitting and $\lesssim1~\umu$K rms for the synchrotron emission
estimation at 80~GHz,\footnote{Note that the rms on the total synchrotron
emission estimated at 80~GHz is significantly lower than that of the
synchrotron + free--free emission presented in Fig.~\ref{fig:spectra_disp}.}
the rms of the polarized synchrotron emission is expected to be
$\lesssim10$~nK. This assumes a median intrinsic fractional polarization of 0.1
in regions of low Galactic latitudes $|b|<30^\circ$. At higher $|b|$, the
polarization fraction is expected to be higher, but synchrotron brightness and
thereby the rms of its estimation is expected to be lower. We therefore expect
that, applying Stokes $Q,U$ fitting to data including the new broadband
spectro-polarimetric measurements at S-Band would allow us to recover the
polarized synchrotron foreground with rms better than $10$~nK per 1 $\deg^2$,
over the entire Southern sky.

Further, the \mmska\ will be sensitive to point sources with flux densities
$\gtrsim$300 mJy beam$^{-1}$ at 2.5~GHz. As it will take about 20 survey cycles
of about 300 hours each to reach the target sensitivity of $\sim$\,0.3~mK in
polarization, as a by-product \mmska\ will allow time-domain study of bright
sources in all four Stokes parameters over 200 frequency channels in the
S-Band.

Also, a Northern-sky counterpart of a similar S-Band survey is financially
viable by procuring one of the SKA-MID dishes when they go into mass production
and installing it in a RFI-quiet region in the Northern hemisphere.
Availability of a Ku-band receiver covering the 12--18~GHz frequency range
provides the possibility to extend QUIJOTE-like foreground studies to the
Southern hemisphere. Thus, the \mmska\ will have strong impact on CMB
foreground science addressing complimentary scientific aspects and help to
expand existing dedicated surveys.  

To summarize, a broad-band spectro-polarimetric survey in S-Band of the
Southern sky with the \mmska\ will be crucial for estimating the synchrotron
foreground, both total and polarized intensities, at frequencies where the CMB
emission is expected to be relatively stronger. The survey with the \mmska\
will be sensitive to angular scales $>1^\circ$ and will be complimentary to
current on-going foreground measurement experiments. Broad-band
spectro-polarimetry at S-Band will provide us the opportunity to apply, for the
first time in studies of the Galactic foreground measurements, the powerful
technique of Stokes $Q,U$ fitting, and constrain the synchrotron + free--free
emission from the radio continuum observations alone. The technique of Stokes
$Q, U$ fitting is expected to play a crucial role in recovering the foreground
polarized signal, both amplitude and polarization angle, at higher frequencies
like 80--100 GHz, where next-generation, space-based CMB polarization
anisotropy studies will be conducted, e.g., the \textit{LiteBIRD} mission.
These make the \mmska\,to have lasting and impactful scientific legacy.

\section*{Acknowledgements}

We thank the anonymous referee for critical and constructive comments that have
improved the content of the paper. We thank Eiichiro Komatsu and Olaf Wucknitz
for valuable comments that have improved the presentation of this paper, and
Pavel Naselsky and Carlo Baccigalupi for interesting discussions. We
acknowledge financial support by the German Federal Ministry of Education and
Research (BMBF) under grant 05A17PB1 (Verbundprojekt D-MeerKAT). BB acknowledges
the generous support of the Simons Foundation. AB would like to thank Deutsche
Bahn for its WIFIonICE and ICEportal services that enabled a significant
portion of the text being written on-board than intended.

\bibliographystyle{mnras}

\bibliography{abasu_etal_cmb.bbl} 

\appendix

\section{Large sky-area broad-bandwidth surveys} \label{sec:survey}

Several radio frequency surveys covering large sky areas have been performed in
the past using narrow frequency bands, mainly to study the Galactic radio
continuum and polarized emission, and hence are not optimum to achieve required
sensitivities for CMB foreground separation {(see
Section~\ref{sec:syncsurvey})}. Further, following the results of previous
dedicated CMB foreground measurement surveys by the {\it WMAP} satellite at
frequencies above $22.8$ GHz and by the {\it Planck} satellite at frequencies
above $28.4$ GHz, it has become abundantly clear that reliable measurements of
the Galactic synchrotron emission are best performed at frequencies between 2
and 20 GHz \citep{planck2015X}.

In this section we discuss in brief the salient features of recent and on-going
efforts to measure the polarized Galactic foreground emissions.

\subsection{C-BASS} \label{sec:cbass}

The C-Band All Sky Survey (C-BASS) is a survey aimed at mapping the entire sky,
both in total and polarized intensities, at 5 GHz (C-Band) with 45 arcmin
angular resolution \citep{jones18}. To cover the entire sky, C-BASS is being
performed with two telescopes located in North and South hemispheres. In the
North, a 6.1-m Gregorian telescope will map the sky for declinations
$>-15^\circ.6$. This telescope uses a receiver at a frequency centred at 4.783
GHz averaging signal over a bandwidth of 0.499 GHz. In the South, a
\mbox{7.6-m} Cassegrain telescope will survey the sky for declination
$<28^\circ.6$ and is equipped with a receiver capable of recording 1 GHz
bandwidth split over 128 frequency channels centred at 5 GHz. To our knowledge,
the C-BASS Southern survey is the first dedicated survey for measuring the
foreground polarized emission contaminating the CMB through broad-bandwidth
radio spectro-polarimetry.

\subsection{S-PASS} \label{sec:spass}

The S-Band Polarization All Sky Survey (S-PASS) is the only other
spectro-polarimetric survey at S-Band. It was performed using the Parkes 64-m
telescope at 9 arcmin angular resolution \citep{carre19}. The survey was
performed using a relatively narrow bandwidth of 256 MHz centred at 2.3 GHz
with 512 frequency channels binned into 8-MHz wide channels. Due to radio
frequency interference, effectively 184 MHz of the band was usable in the
frequency ranges 2.176 to 2.216~GHz and 2.256 to 2.4~GHz. This makes direct
fitting of the Stokes $Q$, $U$ data difficult due to degeneracy between fitting
models and their parameters, {especially in regions of strong Faraday
depolarization}. Also, application of RM-synthesis, even when combined with the
C-BASS Southern sky data, will give rise to strong sidelobes in the Faraday
depth spectrum due to a large gap in $\lambda^2$-space (see
Fig.~\ref{fig:rmsf}). This will make it difficult to discern complicated
Faraday depth features and/or multiple Faraday depth components.

\subsection{QUIJOTE} \label{sec:quijote}

The Q-U-I-JOint TEnerife (QUIJOTE) is an experiment designed to measure
polarized foregrounds to constrain the $B$-mode polarization of CMB in the
Northern sky \citep{quijote15a, quijote15b}. A multi-frequency instrument (MFI)
covers the frequency range 10--20 GHz centred at 11.2, 12.9, 16.7 and 18.7 GHz
with 4 horns. The angular resolution at the two lower frequencies is
$0.92^\circ$ and at the two higher frequencies is $0.60^\circ$. At these
frequencies the foreground continuum emission is expected to be dominated by
the anomalous microwave emission (AME) making QUIJOTE less suitable for
constraining the Galactic synchrotron emission. Moreover, the deep survey of
QUIJOTE for cosmological studies is designed to target smaller sky area
covering around $3000$ deg$^2$. Thus, QUIJOTE will be important to glean
complimentary information on foreground AME emission in contrast to C-BASS and
our planned survey at S-Band.

\subsection{GreenPol} \label{sec:greenpol}

The GreenPol experiment is another effort for measuring the Galactic polarized
emission at frequencies 10, 15, 20, 30 and 44 GHz (Fuskeland et al., in
preparation).\footnote{\url{ https://www.deepspace.ucsb.edu/projects/greenpol}}
Located at the Summit Station in Greenland, the GreenPol can survey upto
$\sim50\%$ of the sky with angular resolutions ranging from 80 arcmin at the
lowest frequency to $\approx18$ arcmin at the highest frequency. Similar to
QUIJOTE, GreenPol {experiment} will be sensitive for studying AME.

\subsection{Surveys below 2 GHz}

The GALFA Continuum Transit
Survey\footnote{\url{https://www.ucalgary.ca/ras/GALFACTS}} (GALFACTS) and the
Global Magneto-Ionic Medium Survey \citep[GMIMS;][]{wolle09} are two large
sky-area, spectro-polarimetric surveys conducted at frequencies below 2 GHz.
GALFACTS is performed with the 300-m Arecibo telescope covering the frequency
range 1.225 to 1.525 GHz. The 300 MHz wide bandwidth is split into 1024
channels. Since, GALFACTS is a transit survey, it covers limited sky area which
is not conducive for CMB foreground studies. GMIMS, however, will cover the
entire sky in the frequency range 0.3 to 1.8 GHz subdivided into low-band,
covering 0.3 to 0.9~GHz, mid-band, covering 0.9 to 1.3~GHz, and high-band,
covering 1.3 to 1.8 GHz. In the high-band, observations are performed with the
DRAO 26-m telescope in the north and Parkes 64-m telescope in the
South.\footnote{The Southern-sky component of GMIMS is known as the Southern
Twenty-cm All-sky Polarization Survey (STAPS).} The low-band survey for the
Southern sky {was performed} using the Parkes 64-m telescope. The
mid-band, 0.9 to 1.3 GHz, is not yet covered.

Although, at these frequencies the Faraday depths can be estimated with better
accuracy as compared to higher frequency measurements, Faraday depolarization
effects are severe. Typically, depolarization increases as $\lambda^4$ making
the polarized signals at $\sim1.5$ GHz $\gtrsim16$ times more depolarized than
at $\sim3$ GHz which results in lower frequency observations being sensitive to
local polarized structures \citep{jelic15}. This is a major concern when
combining the polarization data from frequencies $\lesssim1$ GHz with that from
higher frequencies ($\gtrsim 5$ GHz, e.g. the C-BASS data), because the
emitting volume probed at the two frequency regimes could be significantly
different.

\bsp

\label{lastpage}

\end{document}